\documentclass[11pt,a4paper]{article}

\usepackage{amsmath,scalefnt}
\usepackage{amssymb}
\usepackage{amstext}
\usepackage{latexsym}
\usepackage{graphics}
\usepackage{epsf}
\usepackage{epsfig}
\usepackage{subfigure}
\usepackage{rotating}
\usepackage{curves}
\usepackage{epic}
\usepackage{amsthm}
\usepackage{url}

\usepackage{framed}
\usepackage{adjustbox}
\usepackage{float}
\restylefloat{table}
\usepackage{amsfonts}
\usepackage[T1]{fontenc}
\usepackage[latin1]{inputenc}
\usepackage{authblk}
\usepackage{tikz}
\usetikzlibrary{shapes,arrows}
\usetikzlibrary{decorations.markings}
\usepackage{graphicx}
\usepackage{verbatim}
\usepackage{color}
\usepackage{hyperref}
\usepackage{mathtools}
\usepackage{epstopdf}
\allowdisplaybreaks[1]

\newcommand{\be}{\begin{equation}}
\newcommand{\en}{\end{equation}}

\newtheorem{thm}{Theorem}
\newtheorem{cor}[thm]{Corollary}
\newtheorem{prop}[thm]{Proposition}

\newtheorem{defi}{Definition}[section]
\newtheorem{lem}[defi]{Lemma}

\newtheorem{Theo}{Theorem}[section]

\theoremstyle{remark}
\newtheorem{remark}{\bf{Remark}}[section]
\theoremstyle{remark}

\newcommand{\bedefin}{\begin{defi}}
\newcommand{\findefi}{\end{defi} \medskip}

\newcommand{\betheo}{\begin{theorem}$\!\!${\bf \,\,\,}}
\newcommand{\entheo}{\end{theorem}}
\newcommand{\enth}{\end{theorem}}

\newcommand{\becor}{\begin{cor}$\!\!${\bf .}}
\newcommand{\encor}{\end{cor}}

\newcommand{\belem}{\begin{lem}$\!\!${\bf }}
\newcommand{\enlem}{\end{lem}}

\newcommand{\prf}{\noindent{\bf{ Proof}\,\,}}

\newcommand{\bea}{\begin{eqnarray}}
\newcommand{\ena}{\end{eqnarray}}

\newcommand{\beano}{\begin{eqnarray*}}
\newcommand{\enano}{\end{eqnarray*}}

\newcommand{\bee}{\begin{enumerate}}
\newcommand{\ene}{\end{enumerate}}

\newcommand{\bei}{\begin{itemize}}
\newcommand{\eni}{\end{itemize}}

\newcommand{\betab}{\begin{tabular}}
\newcommand{\entab}{\end{tabular}}

\newcommand{\bd}{\begin{displaymath}}

\DeclareMathOperator{\tr}{Tr}

\catcode `\@=11 \@addtoreset{equation}{section}

\catcode `\@=12

\begin{document}

\title{On Goldman Bracket for $G_2$ Gauge Group }
\author[1,2]{S. Hasibul Hassan Chowdhury\thanks{shhchowdhury@gmail.com}}
\affil[1]{Chern Institute of Mathematics, Nankai University, Tianjin 300071, P. R. China}
\affil[2]{Department of Mathematics and Statistics, Concordia University, Montr\'eal, Qu\'ebec, Canada H3G 1M8}
\date{\today}

\maketitle

{\begin{abstract}
In this paper, we obtain an infinite dimensional Lie algebra of exotic gauge invariant observables that is closed under Goldman-type bracket associated with monodromy matrices of flat connections on a compact Riemann surface for $G_{2}$ gauge group. As a by-product, we give an alternative derivation of known Goldman bracket for classical gauge groups $GL(n,\mathbb{R})$, $SL(n,\mathbb{R})$, $U(n)$, $SU(n)$, $Sp(2n,\mathbb{R})$ and $SO(n)$.
\end{abstract}


}

\section{Introduction}\label{sec:intro}

Traces of monodromy matrices of flat connections computed along closed loops on a Riemann surface $\Sigma$ are known to satisfy the Goldman's Poisson bracket \cite{goldmant} for the following list of gauge groups: $GL(n,\mathbb{R})$, $GL(n,\mathbb{C})$, $GL(n,\mathbb{H})$, $SL(n,\mathbb{R})$, $SL(n,\mathbb{C})$, $SL(n,\mathbb{H})$, $O(p,q)$, $O(n,\mathbb{C})$, $Sp(2n,\mathbb{H})$, $U(p,q)$, $Sp(2n,\mathbb{R})$, $Sp(p,q)$ $Sp(2n,\mathbb{C})$ and $SU(p,q)$. The problem of computation of the bracket between traces of monodromy matrices for exceptional gauge groups so far remains open. The goal of this paper is to fill this gap for $G_{2}$, the simplest exceptional Lie group. In the process, we learn that traces of $G_2$ monodromy matrices are not sufficient to close an algebra under such Poisson bracket and hence demands for the introduction of an infinite set of exotic $G_2$-gauge invariant observables that together with the canonical observables of traces of $G_2$-monodromy matrices form a closed infinite dimensional Lie algebra under such Goldman-type bracket.

Given the fundamental group $\pi$ of a closed oriented surface $S$ and a Lie group $G$, the space $\hbox{Hom}(\pi,G)/G$ is defined as the quotient of the analytic variety $\hbox{Hom}(\pi,G)$ by the action of $G$ by conjugation. In his study (see, for example, \cite{wolpert}) of Weil-Petersson symplectic geometry of Teichm{\"u}ller space (with $G=PSL(2,\mathbb{R})$), Scott A. Wolpert discovered an expression of the Poisson bracket between geodesic length functions (Hamiltonian potential of the flow generated by Fenchel-Nielsen vector fields) in terms of the geometry of the underlying surface. William Goldman, in \cite{goldmant}, went steps further to investigate the geometry of the symplectic structure of $\hbox{Hom}(\pi,G)/G$ for Lie group $G$, satisfying fairly general conditions with the help of family of invariant functions on $\hbox{Hom}(\pi,G)/G$. He obtained the Poisson bracket between such invariant functions due to transversally intersecting homotopically inequivalent loops on $S$. An instructive example of invariant functions in the sense of Goldman \cite{goldmant} comprises of the traces of monodromy matrices of flat connections on the principal $G$-bundle over $S$, also known as {\em Wilson loop} in QCD jargon. M. Chas and D. Sullivan, in the late 90's, generalized Goldman's Poisson bracket by looking at $H_{*}(LM)$, the singular homology of the loop space in a compact oriented manifold $M$ of dimension $d$ (see \cite{chas-sull} for detail). This generalized Poisson bracket, known as the {\em string bracket}, reproduces Goldman's bracket when $d=2$ and $*=0$. All that we have discussed so far is intimately tied with the $\mathbb{Q}$-vector space generated by free homotopy classes of loops on $S$. If this infinite dimensional vector space is quotiented out by null-homotopic loops, then the resulting vector space $V$, equipped with a well-defined bracket $[\;,\;]:V\otimes V\rightarrow V$ induced from the underlying string bracket and a well-defined cobracket $\delta:V\rightarrow V\otimes V$, has a Lie bialgebra structure \cite{turaev}. The cobracket of the Lie bialgebra $(V, [\;,\;], \delta)$ is known as Turaev cobracket in modern literature. Also, of considerable importance in theoretical Physics is the notion of  Batalin-Vilkovisky (BV) operator that arises in the context of {\em string topology} as $\Delta: H_{*}(LM)\rightarrow H_{*+1}(LM)$. Roughly speaking, it produces the $(d+1)$-dimensional family of loops by rotating the $d$-dimensional ones. The string operations make $H_{*}(LM)$ a BV-algebra. Construction of BV-algebra on moduli space of Riemann surfaces and its application in closed string field theory was elucidated by Sen and Zwiebach in \cite{sen-zwiebach}. An excellent treatment of the passage from string topology to {\em topological field theory} by Cattaneo et al. can be found in \cite{cattaneo} where the authors used BV (Lagrangian) or BRST (Hamiltonian) formalism depending on whether the dimension of the base manifold is odd or even, respectively, the relation between the two formalisms being explained in its appendix. Turning ourselves to the physical side, an interesting variant of the usual supersymmetric nonlinear sigma model exhibiting BRST-like fermionic symmetry was studied by Witten in \cite{wittenbrst}.

Consider a $\mathcal{G}$- valued ($\mathcal{G}$ is the Lie algebra of $G$) flat connection $A=A_{z}(z,\bar{z})dz+A_{\bar{z}}(z,\bar{z})d\bar{z}$ on a compact Riemann surface $\Sigma$ of genus $g$. The Atiyah-Bott bracket on the space of flat connections can be derived from the Chern-Simons action on the 3-dimensional manifold $\Sigma\times\mathbb{R}$. Let us represent the connection $1$-forms as $A_{i}=\sum\limits_{a=1}^{n}A^{a}_{i}t_{a}$, where $i=z,\bar{z}$. The generators $\{t_{a}\}$ of the the gauge group $G$ are assumed to satisfy the normalization condition
\begin{equation}\label{intro-orthonormal-basis-con}
\frac{1}{2}\hbox{Tr}(t_{a}t_{b})=f(a)\delta_{ab},
\end{equation}
with $f(a)=\pm 1$.
Then the Atiyah-Bott bracket reads
\begin{equation}\label{intro-AtiyahBott}
\{A_{z}^{a},A_{\bar{z}^{\prime}}^{b}\}=\frac{f(a)}{2}\delta^{ab}\delta^{(2)}(z-z^{\prime}).
\end{equation}
The space of flat connections modulo gauge transformation is finite dimensional and traces of the monodromy matrices of flat connections can be chosen to be the underlying gauge invariant observables.

Goldman in \cite{goldmant} derived the Poisson bracket between traces of the monodromy matrices for classical groups already listed at the start of the introduction. For example, for any two transversally intersecting oriented closed curves $\gamma_{1}$ and $\gamma_{2}$ on $\Sigma$, the Poisson bracket between tarces of $GL(n,\mathbb{R})$ monodromy matrices reads
\begin{equation}\label{intro-eqn-first}
\{\tr M_{\gamma_{1}},\tr M_{\gamma_{2}}\}=\tr M_{\gamma_{1}\circ\gamma_{2}}.
\end{equation}
For the case of $Sp(2n,\mathbb{R})$ the bracket looks as follows
\begin{equation}\label{intro-eqn-second}
\{\tr M_{\gamma_{1}},\tr M_{\gamma_{2}}\}=\frac{1}{2}\left(\tr M_{\gamma_{1}\circ\gamma_{2}}-\tr M_{\gamma_{1}\circ\gamma_{2}^{-1}}\right).
\end{equation}
Here, $\gamma_{1}\circ\gamma_{2}$ and $\gamma_{1}\circ\gamma_{2}^{-1}$ represent loops on $\Sigma$ which are obtained from $\gamma_{1}$ and $\gamma_{2}$ by appropriate resolution of their intersection points (see figure \ref{fig:figure}).

The extensions of Goldman's results to exceptional Lie groups were not known before. Of all five exceptional Lie groups, $G_{2}$ is simultaneously the smallest and one of the most important ones. Recently, it played pivotal roles in exceptional geometry (see \cite{exceptional-holonomy}) and in Lattice QCD (see \cite{excptnl-confinement}, \cite{excptnl-deconfinement}, \cite{g2-gluodynamics} and \cite{color-screening}, for example). Manifolds admitting $G_{2}$ holonomy are also of special interest in M-theory (see \cite{M-theory-g2-holonomy} and \cite{M-theory-g2-holonomy-physics} for a brief review).  Quantization of Goldman bracket for loops on surfaces and its relation to (2+1)-quantum gravity are also investigated in (\cite{quantized-gldmn},\cite{quantized-gldmn-qg}). In (\cite{quantized-gldmn}), it has been shown how signed area phases appear in the quantized version of the classical brackets due to Goldman.

The goal of this paper is to generalize Goldman's bracket to the case of $G_{2}$ gauge group. Consider the $7\times 7$ monodromy matrices $M_{\gamma_{1}}$ and $M_{\gamma_{2}}$ to be in the fundamental representation of $G_{2}$. The existence of a new type of gauge invariant observables emerges by noting that
\begin{equation}\label{intro-eqn-third}
\{\tr M_{\gamma_{1}},\tr M_{\gamma_{2}}\}=\frac{1}{2}\left[\tr M_{\gamma_{1}\circ\gamma_{2}}-\tr M_{\gamma_{1}\circ\gamma_{2}^{-1}}+\frac{1}{3}\sum_{i=1}^{7}\tr(M_{\gamma_{1}}\mathbb{O}_{i})\tr(M_{\gamma_{2}}\mathbb{O}_{i})\right].
\end{equation}
Here $\mathbb{O}_{1},\dots,\mathbb{O}_{7}$ are skew symmetric $7\times 7$ matrices representing the right action of purely imaginary octonions lying in 6-dimensional sphere $S^{6}$ (see \cite{octonionic-rep} for a detailed discussion on left and right octonionic operators). Let $\phi=\sum\limits_{i=1}^{7}\phi_{i}e_{i}$ be a purely imaginary octonion and the matrices $\{\mathbb{O}_{i}\}$ in (\ref{intro-eqn-third}) represent the octonionic imaginary units $\{e_{i}\}$ in the sense of \cite{octonionic-rep}. Then the action of $\mathbb{O}_{i}$ on $\phi$ is defined as
\begin{equation}\label{right-octonionic-oprtr}
\mathbb{O}_{i}\phi=\Im(\phi e_{i}),\qquad i=1,2,\dots,7.
\end{equation}
Therefore, the matrices $M_{\gamma_{1}}\mathbb{O}_{i}$'s and $M_{\gamma_{2}}\mathbb{O}_{i}$'s, appearing in (\ref{intro-eqn-third}), transform the purely imaginary octonions into themselves. A new ingredient of (\ref{intro-eqn-third}) in comparison with the classical Goldman bracket is the term $\sum\limits_{i=1}^{7}\tr(M_{\gamma_{1}}\mathbb{O}_{i})\tr(M_{\gamma_{2}}\mathbb{O}_{i})$. This expression turns out to be gauge invariant although none of the terms $\{\tr(M_{\gamma_{1}}\mathbb{O}_{i})\}$ is individually gauge invariant. We proceed further to show that there is an infinite set of such {\em exotic gauge invariant observables} for the case of $G_2$ gauge group, the Poisson bracket between two such observables being again a linear combination of exotic $G_2$ gauge invariant observables. The Poisson bracket between the trace of a $G_2$ monodromy matrix and an exotic observable of this type can also be found to be a linear combination of exotic  $G_2$- gauge invariant observables, hence proving the closedness of the algebra of $G_2$-gauge invariant observables (canonical+exotic) under such Goldman-type Poisson bracket. 

The organization of the paper is as follows. In section \ref{Goldman-brack-Wilsn}, we recall how the Atiyah-Bott bracket originates from the Hamiltonian Chern-Simons theory and derive an auxiliary expression for Poisson bracket of traces of monodromy matrices along intersecting loops. In section \ref{sec:examples}, we show how this general expression can be used to derive this bracket for a few cases from Goldman's list. In section \ref{sec:G2}, we derive the Poisson bracket between traces of $G_{2}$ monodromy matrices using the formalism developed in section \ref{Goldman-brack-Wilsn} and show how it leads to an infinite set of {\em exotic $G_2$-gauge invariant observables}.

\section{Poisson brackets for traces of monodromy matrices from the Atiyah-Bott bracket}\label{Goldman-brack-Wilsn}
This section is devoted to the review of basic facts of Hamiltonian formulation of Chern-Simons theory which will enable us to understand the symplectic structure of the infinite dimensional phase space of flat gauge connections and eventually lead us to construct the moduli space of flat connections, the dimension of which is given by $(2g-2)\;\hbox{dim}G$ where $g$ is the genus of the underlying Riemann surface $\Sigma$ and $G$ is the gauge group as described in the Introduction \ref{sec:intro}. An elegant treatment of the Hamiltonian formulation of Chern-Simons theory and the related aspects of Goldman's Poisson bracket can be found in \cite{CSref}.

In this paper, space-time is modelled as a $3$-manifold $\Sigma\times\mathbb{R}$ where $\Sigma$ representing \textquotedblleft space\textquotedblright\ is a compact Riemann surface and $\mathbb{R}$ represents \textquotedblleft time\textquotedblright. For an arbitrary real gauge group $G$, the Chern-Simons action functional on $\Sigma\times\mathbb{R}$ reads
\begin{equation}\label{CSaction}
S_{\hbox{\tiny{CS}}}=2\int_{\Sigma\times\mathbb{R}}\tr\;(A\wedge dA+\frac{2}{3}A\wedge A\wedge A).
\end{equation}
The connection $1$-forms on the principal $G$-bundle, taking their values in the Lie algebra $\mathcal{G}$ of the gauge group $G$, are given by
\begin{equation}\label{confrm}
A=A_{z}(z,\bar{z},t)dz+A_{\bar{z}}(z,\bar{z},t)d\bar{z}+A_{0}(z,\bar{z},t)dt.
\end{equation}
Denote the generators of the group $G$ are given by $\{t_{a}\}$ and write
\begin{equation}\label{comps}
A_{i}=\sum_{a=1}^{n}A_{i}^{a}t_{a},
\end{equation}
where the space-time label $i=z,\bar{z},0$. Here, $\{t_{a}\}$ are chosen such that the following holds
\begin{equation}\label{orthonormal-basis-con}
\frac{1}{2}\hbox{Tr}(t_{a}t_{b})=f(a)\delta_{ab},
\end{equation}
with $f(a)=\pm 1$. The curvature form $F=dA+A\wedge A$, for the principal connections (\ref{confrm}), is easily found to vanish. The time component $A_{0}$ of the flat connections can be gauged out. From the gauge fixed Chern-Simons action, one obtains the coordinates and momenta of the underlying phase space of flat connections. At each space-time point, there are $2n$ degrees of freedom associated with $\{A_{z}^{a}\}$ and $\{A_{\bar{z}}^{a}\}$. The Poisson structure of the infinite dimensional space of flat connections is given by the famous Atiyah-Bott bracket between the phase space variables at a given time slice of space-time $\Sigma\times\mathbb{R}$:
\begin{equation}\label{AtiyahBott}
\{A_{z}^{a},A_{\bar{z}^{\prime}}^{b}\}=\frac{f(a)}{2}\delta^{ab}\delta^{(2)}(z-z^{\prime}).
\end{equation}

The space of flat connections modulo gauge transformation turns out to be a finite dimensional space; traces of monodromy matrices of flat connections computed along intersecting loops on $\Sigma$ can be used as gauge invariant observables. In this section, we compute the Poisson bracket between traces of the monodromy matrices along two homotopically inequivalent loops that intersect transversally at a single point using the formalism originating from the Hamiltonian theory of Solitons. The generalization to many intersection points is straight forward.

\begin{figure}[ht]
\begin{framed}
\centering
\subfigure[Two intersecting loops in $\Sigma$]{%
\includegraphics[scale=.145]{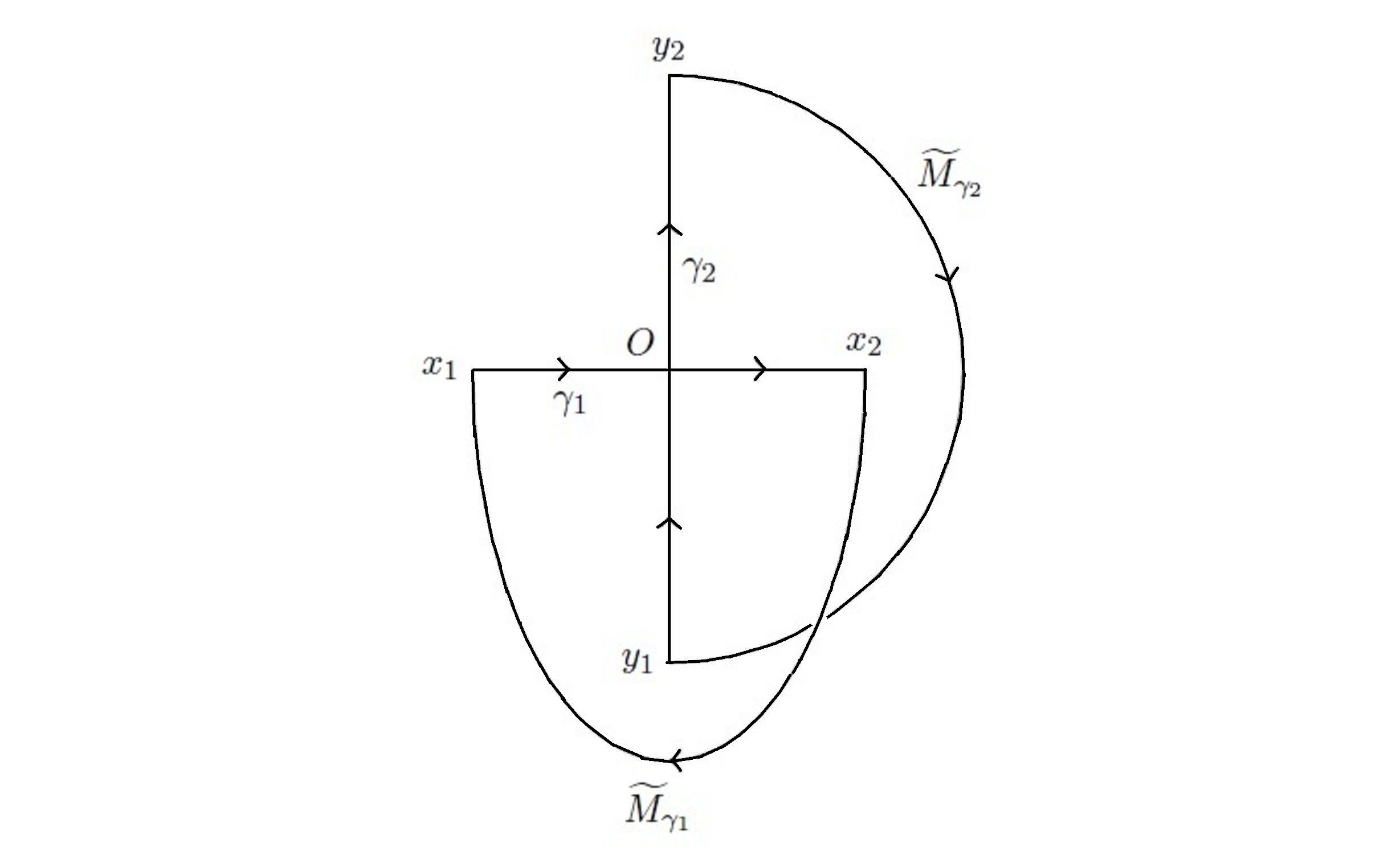}
\label{fig:subfigure1}}
\quad
\subfigure[Superposition $\gamma_{1}\circ\gamma_{2}$]{%
\includegraphics[scale=.16]{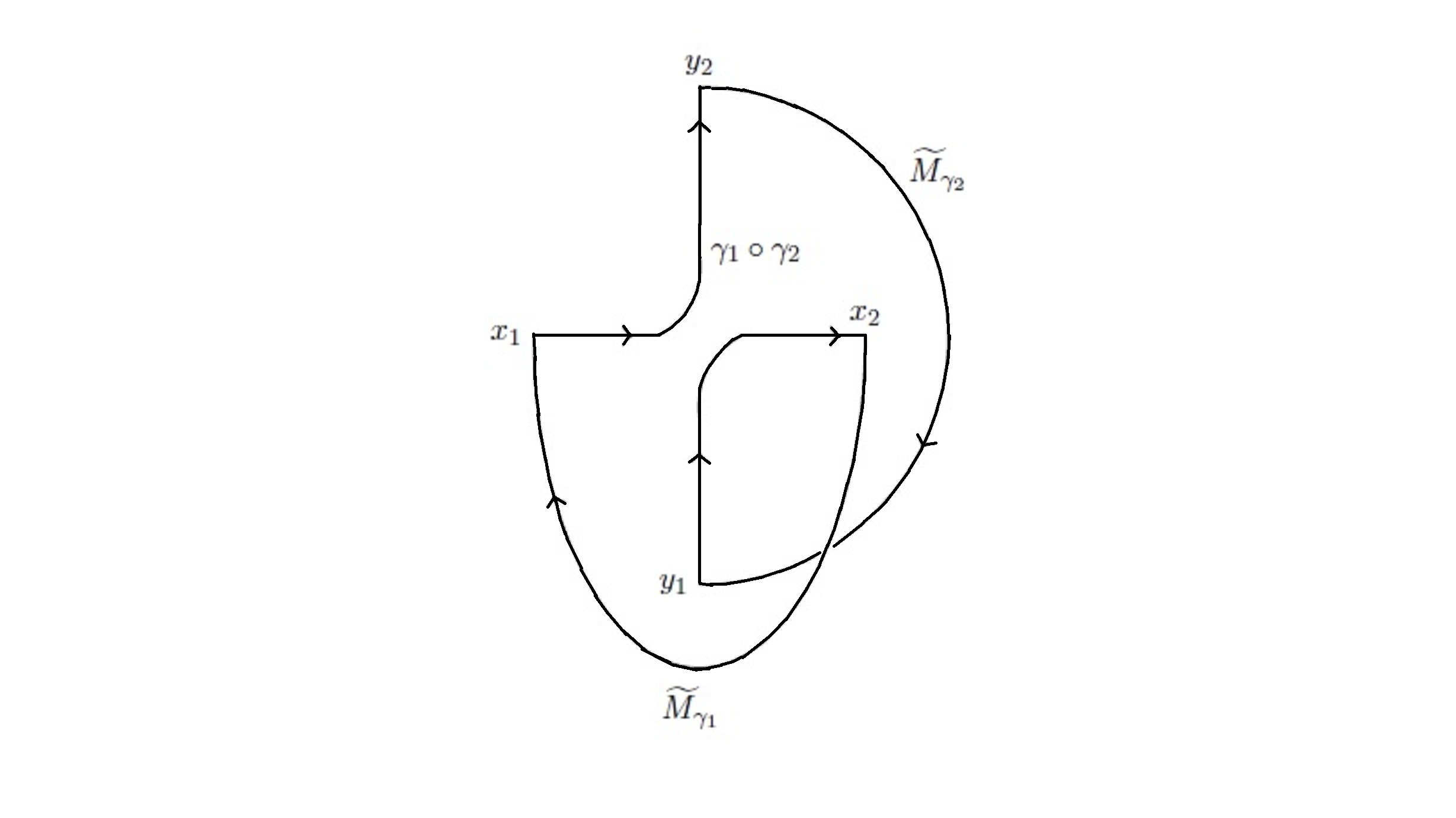}
\label{fig:subfigure2}}
\quad
\subfigure[Superposition $\gamma_{1}\circ\gamma_{2}^{-1}$]{%
\includegraphics[scale=.145]{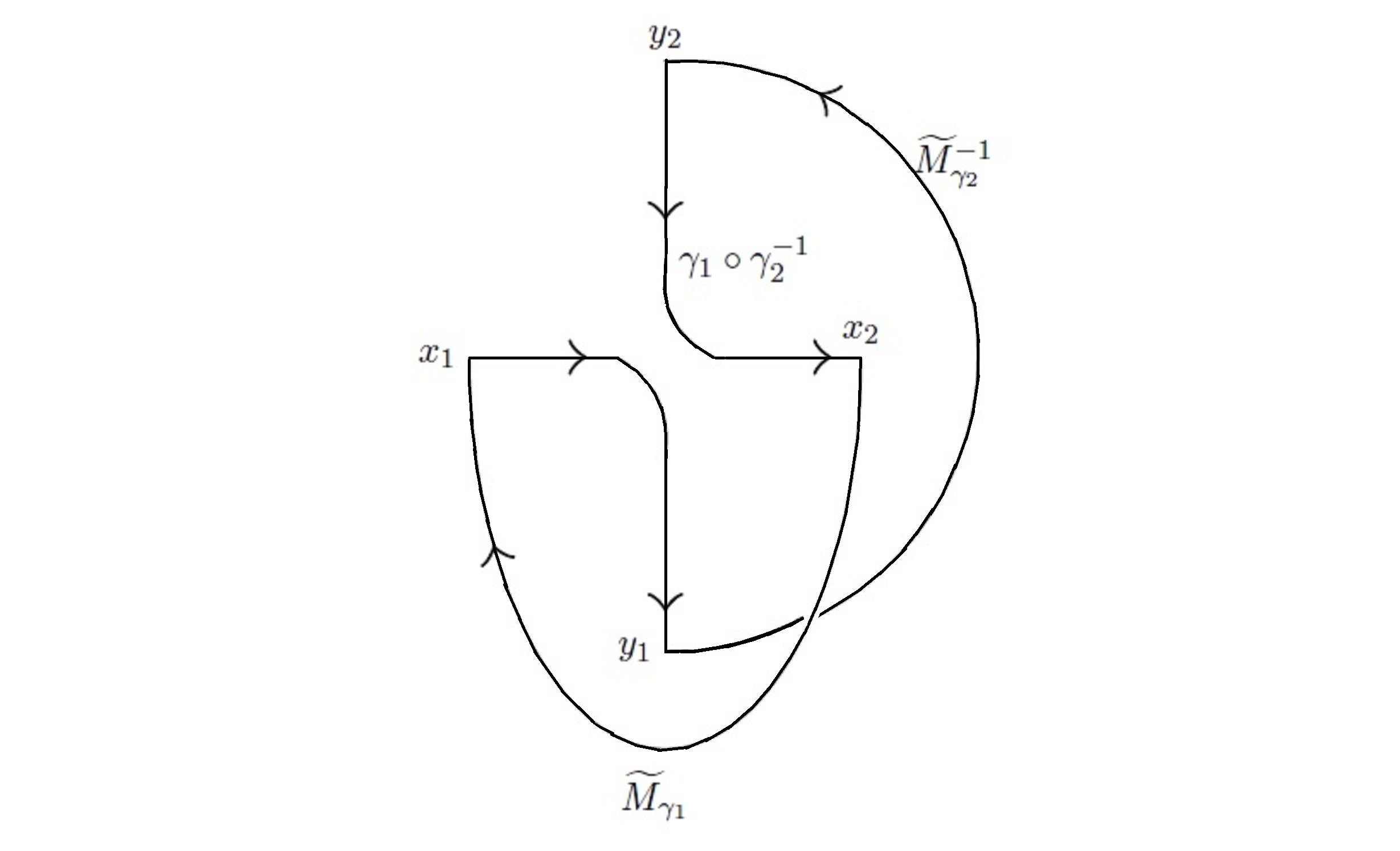}
\label{fig:subfigure3}}
\caption{Transversally intersecting two loops in $\Sigma$ and two different ways of resolving the point of intersection $O$.}
\label{fig:figure}
\end{framed}
\end{figure}

Denote two loops on $\Sigma$ by $\gamma_{1}$ and $\gamma_{2}$ that intersect transversally at a single point on $\Sigma$. Without loss of generality, let us assume that the paths $\gamma_{1}$ and $\gamma_{2}$ intersect orthogonally at $O$. These two loops are illustrated schematically by $x_{1}x_{2}x_{1}$ and $y_{1}y_{2}y_{1}$ in figure \ref{fig:figure}. The parts $x_{1}Ox_{2}$ and $y_{1}Oy_{2}$ are taken to lie along $X$ and $Y$ axes, respectively. The relevant transition matrices are denoted by $T(x_{1},x_{2})$ and $T(y_{1},y_{2})$. Let us denote the monodromy matrices computed along $\gamma_{1}$ and $\gamma_{2}$ by $M_{\gamma_{1}}$ and $M_{\gamma_{2}}$, respectively. They are given by
\begin{equation}\label{Mondrmtot}
\begin{aligned}
M_{\gamma_{1}}&=T(x_{1},x_{2})\widetilde{M}_{\gamma_{1}},\\
M_{\gamma_{2}}&=T(y_{1},y_{2})\widetilde{M}_{\gamma_{2}},
\end{aligned}
\end{equation}
where $\widetilde{M}_{\gamma_{1}}$ and $\widetilde{M}_{\gamma_{2}}$ are the remaining contributions of monodromy matrices $M_{\gamma_{1}}$ and $M_{\gamma_{2}}$ due to the paths $x_{2}x_{1}$ and $y_{2}y_{1}$, respectively (see figure \ref{fig:figure}). The matrices $\widetilde{M}_{\gamma_{1}}$ and $\widetilde{M}_{\gamma_{2}}$ Poisson commute with each other and with other transition matrices in question since they are due to parts of the loops far away from the intersection point $O$ and hence have nothing to do with each other. There are two distinct ways of resolving the point of intersection $O$. One of them is shown in figure \ref{fig:figure} to obtain the loop $\gamma_{1}\circ\gamma_{2}$. Monodromy matrix around the loop $\gamma_{1}\circ\gamma_{2}$ is denoted by $M_{\gamma_{1}\circ\gamma_{2}}$.

Here, the matrices $M_{\gamma_{1}}$ and $M_{\gamma_{2}}$ take their values in the gauge group $G$. Let us represent the connection 1-form $A$ on $\Sigma$ as
\begin{equation}\label{con-onefrm}
A=A_{z}(z,\bar{z})dz+A_{\bar{z}}(z,\bar{z})d\bar{z}=A_{1}(x,y)dx+A_{2}(x,y)dy.
\end{equation}
In view of (\ref{con-onefrm}), the 1-forms, restricted to the real and imaginary axes, read
\begin{equation}\label{restrcetd-form}
\begin{aligned}
&A(x,0)=A_{1}(x,0)dx,\quad\hbox{and}\\
&A(0,y)=A_{2}(0,y)dy.
\end{aligned}
\end{equation}
In terms of the real and imaginary parts of the connection 1-forms, i.e. $A_{1}$ and $A_{2}$, the Atiyah-Bott bracket (\ref{AtiyahBott}) reduces to
\begin{equation}\label{reImAtiyah-Bott}
\{A_{1}^{a}(x,y),A_{2}^{b}(x^{\prime},y^{\prime})\}=\frac{1}{2}f(a)\delta^{ab}\delta(x-x^{\prime})\delta(y-y^{\prime}).
\end{equation}

\begin{lem}\label{fundmntlbracktglc}
The fundamental Poisson brackets between $\mathcal{G}$ valued 1-forms are given by
\begin{equation}\label{eqfundbrackglc}
\{A_{1}(x,0)\overset{\bigotimes}{,} A_{2}(0,y)\}=\frac{1}{2}\delta(x)\delta(y)\Gamma,
\end{equation}
where $\Gamma$ is the Casimir tensor for $\mathcal{G}$ given by
\begin{equation}\label{casimir-tensor}
\Gamma=\sum_{a=1}^{n}f(a)(t_{a}\otimes t_{a}).
\end{equation}
\end{lem}
\prf{.}
The above lemma is just a consequence of (\ref{reImAtiyah-Bott}).
\begin{eqnarray*}
\{A_{1}(x,0)\overset{\otimes}{,}A_{2}(0,y)\}&=& \left\{\sum_{a=1}^{n}A_{1}^{a}(x,0)t_{a}\overset{\bigotimes}{,}\sum_{b=1}^{n}A_{2}^{b}(0,y)t_{b}\right\}\\
&=&\sum_{a=1}^{n}\sum_{b=1}^{n}\{A_{1}^{a}(x,0),A_{2}^{b}(0,y)\}(t_{a}\otimes t_{b})\\
&=&\frac{1}{2}\delta(x)\delta(y)\sum_{a=1}^{n}f(a)(t_{a}\otimes t_{a}).
\end{eqnarray*}
\qed
\begin{remark}
We should emphasize in the context of lemma \ref{fundmntlbracktglc} that the basis of the underlying Lie algebra is chosen in such a way that the trace form between the group generators is diagonalised in order to comply with what was used in the derivation of the Atiyah-Bott bracket (\ref{AtiyahBott}). The statement of lemma \ref{fundmntlbracktglc} is independent of the representation of the Lie algebra, though. All it means is that the same representation has to be chosen during both the derivations of the Atiyah-Bott brackets and the fundamental Poisson brackets.
\end{remark}
Using lemma \ref{fundmntlbracktglc}, one obtains the Poisson bracket between transition matrices along two small paths of the given loops around the intersection point $O$ as illustrated in figure\ref{fig:figure}.
\begin{lem}\label{localtransmtrx}
Let $T(x_{1},x_{2})$ and $T(y_{1},y_{2})$ be the transition matrices corresponding to paths $x_{1}Ox_{2}$ and $y_{1}Oy_{2}$ as indicated in figure \ref{fig:figure}. The Poisson brackets between them is given by
\begin{equation}\label{locltransmtrxeq}
\{T(x_{1},x_{2})\overset{\bigotimes}{,}T(y_{1},y_{2})\}=\frac{1}{2}\left[T(x_{1},0)\otimes T(y_{1},0)\right]\Gamma\left[T(0,x_{2})\otimes T(0,y_{2})\right],
\end{equation}
where $\Gamma$ is the Casimir tensor given by (\ref{casimir-tensor}). Here, $T(x_{1},y_{2})$ and $T(y_{1},x_{2})$, appearing in the right side of (\ref{locltransmtrxeq}), are computed along the loop $\gamma_{1}\circ\gamma_{2}$ of figure \ref{fig:figure}.
\end{lem}
\prf{.}
The Poisson brackets between transition matrices in the context of Hamiltonian theory of Solitons are given in (\cite{faddeevetalt}, page 192). In our setting, this formula gives
\begin{eqnarray}\label{dervtnloctansmtrx}
\lefteqn{\{T(x_{1},x_{2})\overset{\bigotimes}{,}T(y_{1},y_{2})\}}\nonumber\\
&&=\int_{x_{1}}^{x_{2}}\!\!\int_{y_{1}}^{y_{2}}\!\!\!\left[(T(x_{1},x))\otimes T(y_{1},y)\right]\{A_{1}(x,0)\overset{\bigotimes}{,} A_{2}(0,y)\}\left[(T(x,x_{2}))\otimes T(y,y_{2})\right]dx\;dy\nonumber\\
&&=\frac{1}{2}\left[T(x_{1},0)\otimes T(y_{1},0)\right]\Gamma\left[T(0,x_{2})\otimes T(0,y_{2})\right].
\end{eqnarray}
\qed

\begin{lem}\label{Gldmn-brack}
The Poisson bracket between traces of monodromy matrices is as follows
\begin{equation}\label{eqnGldmn-brckt}
\{\tr M_{\gamma_{1}},\tr M_{\gamma_{2}}\}=\frac{1}{2}\tr_{12}[(T(0,x_{2})\widetilde{M}_{\gamma_{1}}T(x_{1},0)\otimes T(0,y_{2})\widetilde{M}_{\gamma_{2}}T(y_{1},0))\Gamma],
\end{equation}
where $M_{\gamma_{1}}$ and $M_{\gamma_{2}}$ are given by (\ref{Mondrmtot}). In (\ref{eqnGldmn-brckt}), $\tr$ and $\tr_{12}$ denote trace in the vector spaces $\mathbb{R}^{n}$ and $\mathbb{R}^{n}\otimes\mathbb{R}^{n}$, respectively.
\end{lem}

\prf{.}
Using (\ref{Mondrmtot}), one obtains
\begin{eqnarray}
\{M_{\gamma_{1}}\overset{\bigotimes}{,}M_{\gamma_{2}}\}&=&\{T(x_{1},x_{2})\widetilde{M}_{\gamma_{1}}\overset{\bigotimes}{,}T(y_{1},y_{2})\widetilde{M}_{\gamma_{2}}\}\nonumber\\
&=&\{T(x_{1},x_{2})\overset{\bigotimes}{,}T(y_{1},y_{2})\widetilde{M}_{\gamma_{2}}\}(\widetilde{M}_{\gamma_{1}}\otimes I_{2})\nonumber\\
&\quad&+[T(x_{1},x_{2})\otimes I_{2}]\{\widetilde{M}_{\gamma_{1}}\overset{\bigotimes}{,}T(y_{1},y_{2})\widetilde{M}_{\gamma_{2}}\}\nonumber\\
&=&\{T(x_{1},x_{2})\overset{\bigotimes}{,}T(y_{1},y_{2})\}(I_{2}\otimes\widetilde{M}_{\gamma_{2}})(\widetilde{M}_{\gamma_{1}}\otimes I_{2}),
\end{eqnarray}
where we have exploited the fact that $\widetilde{M}_{\gamma_{1}}$ and $\widetilde{M}_{\gamma_{2}}$ both Poisson commute with $T(x_{1},x_{2})$ and $T(y_{1},y_{2})$, and among themselves.
Using lemma \ref{localtransmtrx}, one obtains
\begin{eqnarray}\label{finlmondrmcomptn}
\lefteqn{\{M_{\gamma_{1}}\overset{\bigotimes}{,}M_{\gamma_{2}}\}}\nonumber\\
&&=\{T(x_{1},x_{2})\overset{\bigotimes}{,}T(y_{1},y_{2})\}(\widetilde{M}_{\gamma_{1}}\otimes\widetilde{M}_{\gamma_{2}})\nonumber\\
&&=\frac{1}{2}\left[T(x_{1},0)\otimes T(y_{1},0)\right]\Gamma\left[T(0,x_{2})\otimes T(0,y_{2})\right](\widetilde{M}_{\gamma_{1}}\otimes\widetilde{M}_{\gamma_{2}})\nonumber\\
&&=\frac{1}{2}\left[T(x_{1},0)\otimes T(y_{1},0)\right]\Gamma[T(0,x_{2})\widetilde{M}_{\gamma_{1}}\otimes T(0,y_{2})\widetilde{M}_{\gamma_{2}}].
\end{eqnarray}

Taking trace on both sides of equation (\ref{finlmondrmcomptn}) and subsequently making use of the cyclic property of trace, one finally obtains
\begin{equation}\label{final-exprssn}
\{\tr M_{\gamma_{1}},\tr M_{\gamma_{2}}\}=\frac{1}{2}\tr_{12}[(T(0,x_{2})\widetilde{M}_{\gamma_{1}}T(x_{1},0)\otimes T(0,y_{2})\widetilde{M}_{\gamma_{2}}T(y_{1},0))\Gamma].
\end{equation}
\qed

\section{Examples of Poisson brackets between traces of monodromy matrices for some known real Lie groups}\label{sec:examples}
In the previous section, we obtained an auxiliary formula (\ref{eqnGldmn-brckt}) for Poisson brackets between traces of monodromy matrices computed along free homotopy classes of loops on $\Sigma$. In this section, we shall use it to reproduce Goldman's brackets for $GL(n,\mathbb{R})$, $U(n)$, $SL(n,\mathbb{R})$, $SU(n)$, $Sp(2n,\mathbb{R})$ and $SO(n)$ gauge groups.

We, first, note that the generalized Gell-Mann matrices in $n$ dimensions read
\begin{eqnarray}\label{gell-mann-matrices}
\begin{aligned}
&h^{n}_{1}=\sqrt{\frac{2}{n}}\sum_{i=1}^{n}e_{ii},\\
&h^{n}_{k}=\sqrt{\frac{2}{k(k-1)}}\sum_{i=1}^{k-1}e_{ii}-\sqrt{2-\frac{2}{k}}e_{kk},\quad\hbox{for $1<k\leq n$},\\
&f^{n}_{k,j}=e_{kj}+e_{jk},\quad\hbox{for $k<j$},\\
&f^{n}_{k,j}=-i(e_{jk}-e_{kj}),\quad\hbox{for $k>j$}.
\end{aligned}
\end{eqnarray}
Here, $e_{jk}$ is an $n\times n$ matrix with $1$ in the $(j,k)$ entry and $0$ elsewhere.

A couple of preparatory lemmas are required in order to derive the Poisson bracket between traces of monodromy matrices for some known real Lie groups from (\ref{eqnGldmn-brckt}).
\begin{lem}\label{tensor-product-formula1}
Given the matrices $h^{n}_{1}$ and $h^{n}_{k}$ as in (\ref{gell-mann-matrices}), we have
\begin{equation}\label{fomula-1}
h^{n}_{1}\otimes h^{n}_{1}+\sum_{k=2}^{n}h^{n}_{k}\otimes h^{n}_{k}=2e_{11}\otimes e_{11}+2\sum_{k=2}^{n}e_{kk}\otimes e_{kk}.
\end{equation}
\end{lem}
\prf{.}
\begin{eqnarray}
\lefteqn{\sum_{k=2}^{n}h^{n}_{k}\otimes h^{n}_{k}}\nonumber\\
&&=\sum_{k=2}^{n}\left[\left(\sqrt{\frac{2}{k(k-1)}}\sum_{i=1}^{k-1}e_{ii}-\sqrt{2-\frac{2}{k}}e_{kk}\right)\otimes\left(\sqrt{\frac{2}{k(k-1)}}\sum_{i=1}^{k-1}e_{ii}-\sqrt{2-\frac{2}{k}}e_{kk}\right)\right]\nonumber\\
&&=\sum_{k=2}^{n}\left[\frac{2}{k-1}\left(\sum_{i=1}^{k-1}e_{ii}\otimes\sum_{i=1}^{k-1}e_{ii}\right)-\frac{2}{k}\left(\sum_{i=1}^{k-1}e_{ii}\otimes\sum_{i=1}^{k-1}e_{ii}\right)-\frac{2}{k}\sum_{i=1}^{k-1}e_{ii}\otimes e_{kk}\right.\nonumber\\
&&\left.\qquad-\frac{2}{k}\sum_{i=1}^{k-1}e_{kk}\otimes e_{ii}+\left(2-\frac{2}{k}\right)e_{kk}\otimes e_{kk}\right]\label{forml2}.
\end{eqnarray}
Consider the right side of (\ref{forml2}) and compute
\begin{eqnarray}
\lefteqn{\sum_{k=2}^{n}\left[\frac{2}{k-1}\left(\sum_{i=1}^{k-1}e_{ii}\otimes\sum_{i=1}^{k-1}e_{ii}\right)-\frac{2}{k}\left(\sum_{i=1}^{k-1}e_{ii}\otimes\sum_{i=1}^{k-1}e_{ii}\right)\right]}\nonumber\\
&&=\sum_{k=2}^{n}\frac{2}{k-1}\left(\sum_{i=1}^{k-1}e_{ii}\otimes\sum_{i=1}^{k-1}e_{ii}\right)-\sum_{k=3}^{n+1}\frac{2}{k-1}\left(\sum_{i=1}^{k-2}e_{ii}\otimes\sum_{i=1}^{k-2}e_{ii}\right)\nonumber\\
&&=\sum_{k=2}^{n}\frac{2}{k-1}\left(\sum_{i=1}^{k-1}e_{ii}\otimes\sum_{i=1}^{k-1}e_{ii}\right)-\sum_{k=3}^{n+1}\frac{2}{k-1}\left[\left(\sum_{i=1}^{k-1}e_{ii}-e_{k-1,k-1}\right)\right.\nonumber\\
&&\left.\quad\otimes\left(\sum_{i=1}^{k-1}e_{ii}-e_{k-1,k-1}\right)\right]\nonumber\\
&&=\sum_{k=2}^{n}\frac{2}{k-1}\left(\sum_{i=1}^{k-1}e_{ii}\otimes\sum_{i=1}^{k-1}e_{ii}\right)-\sum_{k=3}^{n+1}\frac{2}{k-1}\left(\sum_{i=1}^{k-1}e_{ii}\otimes\sum_{i=1}^{k-1}e_{ii}\right)\nonumber\\
&&+\sum_{k=3}^{n+1}\frac{2}{k-1}\left(\sum_{i=1}^{k-1}e_{ii}\otimes e_{k-1,k-1}\right)+\sum_{k=3}^{n+1}\frac{2}{k-1}\left(\sum_{i=1}^{k-1}e_{k-1,k-1}\otimes e_{ii}\right)\nonumber\\
&&-\sum_{k=3}^{n+1}\frac{2}{k-1}(e_{k-1,k-1}\otimes e_{k-1,k-1})\nonumber\\
&&=2e_{11}\otimes e_{11}-\frac{2}{n}\left(\sum_{i=1}^{n}e_{ii}\otimes\sum_{i=1}^{n}e_{ii}\right)+\sum_{k=3}^{n+1}\frac{2}{k-1}\left(\sum_{i=1}^{k-1}e_{ii}\otimes e_{k-1,k-1}\right)\nonumber\\
&&+\sum_{k=3}^{n+1}\frac{2}{k-1}\left(\sum_{i=1}^{k-1}e_{k-1,k-1}\otimes e_{ii}\right)-\sum_{k=3}^{n+1}\frac{2}{k-1}(e_{k-1,k-1}\otimes e_{k-1,k-1})\nonumber\\
&&=2e_{11}\otimes e_{11}-\frac{2}{n}\left(\sum_{i=1}^{n}e_{ii}\otimes\sum_{i=1}^{n}e_{ii}\right)+\sum_{k=2}^{n}\frac{2}{k}\left(\sum_{i=1}^{k}e_{ii}\otimes e_{kk}\right)\nonumber\\
&&+\sum_{k=2}^{n}\frac{2}{k}\left(\sum_{i=1}^{k}e_{kk}\otimes e_{ii}\right)-\sum_{k=2}^{n}\frac{2}{k}(e_{kk}\otimes e_{kk})\nonumber\\
&&=2e_{11}\otimes e_{11}-\frac{2}{n}\left(\sum_{i=1}^{n}e_{ii}\otimes\sum_{i=1}^{n}e_{ii}\right)+\sum_{k=2}^{n}\frac{2}{k}\left(\sum_{i=1}^{k-1}e_{ii}\otimes e_{kk}\right)+\sum_{k=2}^{n}\frac{2}{k}(e_{kk}\otimes e_{kk})\nonumber\\
&&+\sum_{k=2}^{n}\frac{2}{k}\left(\sum_{i=1}^{k-1}e_{kk}\otimes e_{ii}\right)+\sum_{k=2}^{n}\frac{2}{k}(e_{kk}\otimes e_{kk})-\sum_{k=2}^{n}\frac{2}{k}(e_{kk}\otimes e_{kk})\nonumber\\
&&=2e_{11}\otimes e_{11}-\frac{2}{n}\left(\sum_{i=1}^{n}e_{ii}\otimes\sum_{i=1}^{n}e_{ii}\right)+\sum_{k=2}^{n}\frac{2}{k}\left(\sum_{i=1}^{k-1}e_{ii}\otimes e_{kk}\right)+\sum_{k=2}^{n}\frac{2}{k}(e_{kk}\otimes e_{kk})\nonumber\\
&&+\sum_{k=2}^{n}\frac{2}{k}\left(\sum_{i=1}^{k-1}e_{kk}\otimes e_{ii}\right)\label{forml3}.
\end{eqnarray}
Plugging (\ref{forml3}) into (\ref{forml2}), one obtains
\begin{equation}\label{totl-tensor-formula}
\sum_{k=2}^{n}h^{n}_{k}\otimes h^{n}_{k}=2e_{11}\otimes e_{11}-\frac{2}{n}\left(\sum_{i=1}^{n}e_{ii}\otimes\sum_{i=1}^{n}e_{ii}\right)+2\sum_{k=2}^{n}e_{kk}\otimes e_{kk},
\end{equation}
which leads to
\begin{equation*}
h^{n}_{1}\otimes h^{n}_{1}+\sum_{k=2}^{n}h^{n}_{k}\otimes h^{n}_{k}=2e_{11}\otimes e_{11}+2\sum_{k=2}^{n}e_{kk}\otimes e_{kk}.
\end{equation*}
\qed

\begin{lem}\label{tensor product-formula2}
Given $f^{n}_{k,j}$ for $k<j$ and $k>j$ as in (\ref{gell-mann-matrices}), the following holds
\begin{equation}\label{second-tensr-forml}
\sum_{k\neq j}f^{n}_{k,j}\otimes f^{n}_{k,j}=2\sum_{k\neq j}e_{jk}\otimes e_{kj}.
\end{equation}
\end{lem}
\prf{.}
\begin{eqnarray}\label{greater}
\sum_{k<j}f^{n}_{kj}\otimes f^{n}_{kj}&=&\sum_{k<j}(e_{kj}+e_{jk})\otimes(e_{kj}+e_{jk})\nonumber\\
&=&\sum_{k<j}e_{kj}\otimes e_{kj}+\sum_{k<j}e_{kj}\otimes e_{jk}+\sum_{k<j}e_{jk}\otimes
e_{kj}+\sum_{k<j}e_{jk}\otimes e_{jk}.\nonumber\\
\end{eqnarray}
We also have,
\begin{eqnarray}\label{less}
\sum_{k>j}f^{n}_{kj}\otimes f^{n}_{kj}&=&\sum_{k>j}(-ie_{jk}+ie_{kj})\otimes(-ie_{jk}+ie_{kj})\nonumber\\
&=&-\sum_{k>j}e_{jk}\otimes e_{jk}+\sum_{k>j}e_{jk}\otimes e_{kj}+\sum_{k>j}e_{kj}\otimes
e_{jk}-\sum_{k>j}e_{kj}\otimes e_{kj}.\nonumber\\
\end{eqnarray}
Therefore, (\ref{greater}) together with (\ref{less}) imply
\begin{eqnarray}\label{second-part}
\sum_{k\neq j}f^{n}_{k,j}\otimes f^{n}_{k,j}&=&2\sum_{k>j}e_{jk}\otimes e_{kj}+2\sum_{k<j}e_{jk}\otimes e_{kj}\nonumber\\
&=&2\sum_{k\neq j}e_{jk}\otimes e_{kj}.
\end{eqnarray}
\qed

Let us now compute the Casimir tensor $\Gamma$ appearing in (\ref{eqnGldmn-brckt}) for the specific cases of $GL(n,\mathbb{R})$, $U(n)$, $SL(n,\mathbb{R})$, $SU(n)$ and $SO(n)$. In what follows, the  $n^{2}\times n^{2}$ permutation matrix is denoted by $P$. Given two $n\times n$ matrices $A$ and $B$, $P$ enjoys the following properties:
\begin{equation}\label{properties-permutation}
\begin{split}
&P(A\otimes B)=(B\otimes A)P \\
&\tr_{12}[(A\otimes B)P]=\tr(AB).
\end{split}
\end{equation}

\begin{prop}\label{casimir-first-prop}
For $GL(n,\mathbb{R})$ and $U(n)$ gauge groups, the Casimir tensor in the auxiliary expression (\ref{eqnGldmn-brckt}) of Poisson bracket between traces of the monodromy matrices reads
\begin{equation}\label{casimir-GL-U}
\Gamma=2P,
\end{equation}
where $P=\sum\limits_{k,j=1}^{n} e_{jk}\otimes e_{kj}$ is the Permutation matrix.
\end{prop}
\prf{.}

\textbf{Case $1$: $GL(n,\mathbb{R})$}\\
The Lie algebra associated with $GL(n,\mathbb{R})$ is $\mathfrak{gl}(n,\mathbb{R})$, the vector space of all real $n\times n$ matrices. The dimension of this vector space is $n^{2}$.
We choose the matrix $h_{1}^{n}$, $(n-1)$ matrices $h^{n}_{k}$ with $1<k\leq n$, $\frac{(n^{2}-n)}{2}$ matrices $f^{n}_{k,j}$ with $k<j$, and another $\frac{(n^{2}-n)}{2}$ matrices $if^{n}_{k,j}$ with $k>j$ from (\ref{gell-mann-matrices}) to form a basis of $\mathfrak{gl}(n,\mathbb{R})$. Here, in (\ref{orthonormal-basis-con}), associated with the preceding choice of generators for $GL(n,\mathbb{R})$, $f(a)=-1$ for the $\frac{(n^{2}-n)}{2}$ basis elements $if^{n}_{k,j}$ with $k>j$. For the rest of the $n^{2}$ basis elements, we have $f(a)=1$.

With the above choice of the basis of $\mathfrak{gl}(n,\mathbb{R})$, the Casimir tensor $\Gamma$ reads,
\begin{eqnarray}
\Gamma&=&h^{n}_{1}\otimes h^{n}_{1}+\sum_{k=2}^{n}h^{n}_{k}\otimes h^{n}_{k}+\sum_{k<j}f^{n}_{k,j}\otimes f^{n}_{k,j}+\sum_{k>j}-(if^{n}_{k,j}\otimes if^{n}_{k,j})\nonumber\\
&=&h^{n}_{1}\otimes h^{n}_{1}+\sum_{k=2}^{n}h^{n}_{k}\otimes h^{n}_{k}+\sum_{k<j}f^{n}_{k,j}\otimes f^{n}_{k,j}+\sum_{k>j}(f^{n}_{k,j}\otimes f^{n}_{k,j}).\label{casimir-gl}
\end{eqnarray}
Using lemma \ref{tensor-product-formula1} together with lemma \ref{tensor product-formula2} in (\ref{casimir-gl}), one obtains the Casimir tensor for the case of $GL(n,\mathbb{R})$ gauge group:
\begin{equation}\label{final-gln}
\Gamma=2\sum\limits_{k,j=1}^{n} e_{jk}\otimes e_{kj}.
\end{equation}

\textbf{Case $2$: $U(n)$}\\
An appropriate choice of basis for the Lie algebra $\mathfrak{u}(n)$, in the context of (\ref{orthonormal-basis-con}), will be the $n^{2}$ skew-Hermitian matrices (see (\ref{gell-mann-matrices})) $ih^{n}_{1}$, $ih^{n}_{k}$ for $1<k\leq n$ and $if^{n}_{k,j}$ for $k\neq j$. In accordance with the choice of these generators of unitary group $U(n)$, $f(a)=-1$ in (\ref{orthonormal-basis-con}) for $a=1,2\dots, n^{2}$. The corresponding Casimir tensor $\Gamma$ reads off immediately
\begin{eqnarray}
\Gamma&=&-(ih^{n}_{1}\otimes ih^{n}_{1})+\sum_{k=2}^{n}-(ih^{n}_{k}\otimes ih^{n}_{k})+\sum_{k<j}-(if^{n}_{k,j}\otimes if^{n}_{k,j})+\sum_{k>j}-(if^{n}_{k,j}\otimes if^{n}_{k,j})\nonumber\\
&=&h^{n}_{1}\otimes h^{n}_{1}+\sum_{k=2}^{n}h^{n}_{k}\otimes h^{n}_{k}+\sum_{k<j}f^{n}_{k,j}\otimes f^{n}_{k,j}+\sum_{k>j}f^{n}_{k,j}\otimes f^{n}_{k,j}\nonumber\\
&=&2\sum\limits_{k,j=1}^{n} e_{jk}\otimes e_{kj}.\label{un-proof}
\end{eqnarray}
Here, again, we use lemma \ref{tensor-product-formula1} and lemma \ref{tensor product-formula2} to arrive at (\ref{un-proof}).
\qed

Direct application of proposition \ref{casimir-first-prop} in (\ref{eqnGldmn-brckt}) and subsequent use of the properties of $P$, enumerated in (\ref{properties-permutation}), yield the formula of Poisson bracket for traces of $GL(n,\mathbb{R})$ and $U(n)$ monodromy matrices as given by the following theorem:
\begin{thm}\label{theorem-gln-un}
The Poisson bracket between traces of $GL(n,\mathbb{R})$ or $U(n)$ monodromy matrices reads
\begin{equation}\label{goldman-bracket-gln-sun}
\{\tr M_{\gamma_{1}},\tr M_{\gamma_{2}}\}=\tr M_{\gamma_{1}\circ\gamma_{2}},
\end{equation}
where $M_{\gamma_{1}\circ\gamma_{2}}$ is a $GL(n,\mathbb{R})$ or $U(n)$ monodromy matrix computed along the loop $\gamma_{1}\circ\gamma_{2}$ of figure \ref{fig:figure}.
\end{thm}

\begin{prop}\label{prop2-sln-sun}
The Casimir tensor in (\ref{eqnGldmn-brckt}) for $SL(n,\mathbb{R})$ or $SU(n)$ gauge group reads
\begin{equation}\label{casimir-SL-SU}
\Gamma=2P-\frac{2}{n}\mathbb{I},
\end{equation}
with $P=\sum\limits_{k,j=1}^{n} e_{jk}\otimes e_{kj}$ being the Permutation matrix and $\mathbb{I}$ being the $n^{2}\times n^{2}$ identity matrix.
\end{prop}
\prf{.}

\textbf{Case $1$: $SL(n,\mathbb{R})$}\\
The Lie algebra $\mathfrak{sl}(n,\mathbb{R})$ consists of traceless $n\times n$ real matrices. We, therefore, choose $(n-1)$ matrices $h^{n}_{k}$ with $1<k\leq n$, $\frac{(n^{2}-n)}{2}$ matrices $f^{n}_{kj}$ for $k\leq j$ and another $\frac{(n^{2}-n)}{2}$ real matrices $if^{n}_{kj}$ with $k>j$ from the ones enumerated in (\ref{gell-mann-matrices}). As was in the case of $\mathfrak{gl}(n,\mathbb{R})$, $f(a)=-1$ in (\ref{orthonormal-basis-con}) holds only for the $SL(n,\mathbb{R})$ group generators given by $if^{n}_{kj}$. Therefore, the associated Casimir tensor reads
\begin{eqnarray}
\Gamma&=&\sum_{k=2}^{n}h^{n}_{k}\otimes h^{n}_{k}+\sum_{k<j}f^{n}_{k,j}\otimes f^{n}_{k,j}+\sum_{k>j}-(if^{n}_{k,j}\otimes if^{n}_{k,j})\nonumber\\
&=&\sum_{k=2}^{n}h^{n}_{k}\otimes h^{n}_{k}+\sum_{k<j}f^{n}_{k,j}\otimes f^{n}_{k,j}+\sum_{k>j}f^{n}_{k,j}\otimes f^{n}_{k,j}\nonumber\\
&=&2P-h^{n}_{1}\otimes h^{n}_{1}\nonumber\\
&=&2P-\frac{2}{n}\mathbb{I}.\label{casimir-tensor-sln}
\end{eqnarray}

\textbf{Case $2$: $SU(n)$}\\
The real Lie algebra $\mathfrak{su}(n)$ consists of $n\times n$ traceless skew-Hermitian matrices. As a basis of $\mathfrak{su}(n)$, we choose $(n-1)$ traceless skew-Hermitian matrices $ih^{n}_{k}$ with $1<k\leq n$ and another $(n^{2}-n)$ such matrices $if^{n}_{kj}$ for $k\neq j$ from the matrices enumerated in (\ref{gell-mann-matrices}). Here, we only have $f(a)=-1$ in (\ref{orthonormal-basis-con}) for all such $(n^{2}-1)$ group generators of $SU(n)$. The corresponding Casimir tensor then reads
\begin{eqnarray}
\Gamma&=&\sum_{k=2}^{n}-(ih^{n}_{k}\otimes ih^{n}_{k})+\sum_{k<j}-(if^{n}_{k,j}\otimes if^{n}_{k,j})+\sum_{k>j}-(if^{n}_{k,j}\otimes if^{n}_{k,j})\nonumber\\
&=&\sum_{k=2}^{n}h^{n}_{k}\otimes h^{n}_{k}+\sum_{k<j}f^{n}_{k,j}\otimes f^{n}_{k,j}+\sum_{k>j}f^{n}_{k,j}\otimes f^{n}_{k,j}\nonumber\\
&=&2P-h^{n}_{1}\otimes h^{n}_{1}\nonumber\\
&=&2P-\frac{2}{n}\mathbb{I}.\label{casimir-tensor-sun}
\end{eqnarray}
We have repeatedly used lemma \ref{tensor-product-formula1} and lemma \ref{tensor product-formula2} in establishing (\ref{casimir-tensor-sln}) and (\ref{casimir-tensor-sun}).\\
\qed

Following the use of proposition \ref{prop2-sln-sun} in (\ref{eqnGldmn-brckt}) and subsequent use of the properties of $P$ as given by (\ref{properties-permutation}), one obtains the Poisson bracket for $SL(n,\mathbb{R})$ and $SU(n)$ monodromy matrices.
\begin{thm}\label{theorem-sln-sun}
The Poisson bracket between traces of monodromy matrices for $SL(n,\mathbb{R})$ and $SU(n)$ gauge groups is given by
\begin{equation}\label{goldman-bracket-sln-sun}
\{\tr M_{\gamma_{1}},\tr M_{\gamma_{2}}\}=\tr M_{\gamma_{1}\circ\gamma_{2}}-\frac{1}{n}\tr M_{\gamma_{1}}\tr M_{\gamma_{2}}.
\end{equation}
\end{thm}
In course of proving theorem \ref{theorem-sln-sun}, one also makes use of the identity $\tr_{12}(A\otimes B)=\tr A\tr B$ for any two $n\times n$ matrices $A$ and $B$.

We shall now consider the case when the gauge group is $Sp(2n,\mathbb{R})$. It is being dealt separately since an appropriate choice of basis for the associated Lie algebra $\mathfrak{sp}(2n,\mathbb{R})$, in view of (\ref{orthonormal-basis-con}), is unrelated with the generalized Gell-Mann matrices enumerated in (\ref{gell-mann-matrices}).

The Lie algebra $\mathfrak{sp}(2n,\mathbb{R})$ is an $n(2n+1)$ dimensional real vector space. An appropriate choice of basis, along with respective $f(a)=\pm 1$ for $a=1, 2, \dots, n(2n+1)$ in (\ref{orthonormal-basis-con}), is outlined in the following table:

\begin{table}[H]
\begin{center}
    \begin{tabular}{| l | l | l | l |}
    \hline
    $\bf{i,j,k}$ & \bf{Basis elements} & $\bf{f(a)}$ & \bf{No. of elements}\\ \hline
    $1\leq i< j\leq n$ & $\frac{1}{\sqrt{2}}(e_{i,j+n}+e_{j,i+n}+e_{j+n,i}+e_{i+n,j})$ & 1 & $\frac{n^{2}-n}{2}$\\ \hline
    $1\leq i< j\leq n$ & $\frac{1}{\sqrt{2}}(e_{i,j+n}+e_{j,i+n}-e_{j+n,i}-e_{i+n,j})$    & -1 & $\frac{n^{2}-n}{2}$\\ \hline
    $1\leq k\leq n$ & $e_{k, n+k}+e_{n+k, k}$ & 1 & n \\ \hline
    $1\leq k\leq n$ & $e_{k, n+k}-e_{n+k, k}$ & -1 & n \\ \hline
    $1\leq i< j\leq n$ & $\frac{1}{\sqrt{2}}(e_{ij}+e_{ji}-e_{i+n,j+n}-e_{j+n,i+n})$ & 1 & $\frac{n^{2}-n}{2}$\\ \hline
    $1\leq i< j\leq n$ & $\frac{1}{\sqrt{2}}(e_{ij}-e_{ji}+e_{i+n,j+n}-e_{j+n,i+n})$ & -1 & $\frac{n^{2}-n}{2}$\\ \hline
    $1\leq k\leq n$ & $e_{kk}-e_{k+n, k+n}$ & 1 & n \\ \hline
    \end{tabular}
    \caption{Appropriate choice of basis for $\mathfrak{sp}(2n,\mathbb{R})$}
\end{center}
\end{table}

Now, the Casimir tensor for the structure Lie group $Sp(2n,\mathbb{R})$ is provided by the following proposition
\begin{prop}\label{sp2n-prop}
The Casimir tensor $\Gamma$ in (\ref{eqnGldmn-brckt}), for $Sp(2n,\mathbb{R})$ gauge group, reads
\begin{equation}\label{casimir-sp2n}
\Gamma=P+\chi,
\end{equation}
with $\chi$ given as
\begin{align}\label{defect-matrix}
\chi&=\sum_{1\leq i<j\leq n}(e_{i,j+n}\otimes e_{i+n,j}+e_{j,i+n}\otimes e_{j+n,i}+e_{j+n,i}\otimes e_{j,i+n}+e_{i+n,j}\otimes e_{i,j+n}\nonumber\\
&-e_{ij}\otimes e_{i+n,j+n}-e_{j+n,i+n}\otimes e_{ji}-e_{ji}\otimes e_{j+n,i+n}-e_{i+n,j+n}\otimes e_{ij})\nonumber\\
&+\sum_{1\leq k\leq n}(e_{k,n+k}\otimes e_{n+k,k}+e_{n+k,k}\otimes e_{k,n+k}-e_{kk}\otimes e_{k+n,k+n}-e_{k+n,k+n}\otimes e_{kk}).
\end{align}
We shall be calling $\chi$ as the defect matrix henceforth.
\end{prop}
\prf{.}
In order to prove proposition \ref{sp2n-prop}, we first note that, for any two $n\times n$ matrices $a$ and $b$, the following holds
\begin{equation}\label{helpful-frml}
(a+b)\otimes(a+b)-(a-b)\otimes(a-b)=2(a\otimes b+b\otimes a).
\end{equation}
Using the above fact, we have the following for $1\leq i<j\leq n$:
\begin{eqnarray}\label{prop-spn-proof-first}
&&\frac{1}{\sqrt{2}}(e_{i,j+n}+e_{j,i+n}+e_{j+n,i}+e_{i+n,j})\otimes\frac{1}{\sqrt{2}}(e_{i,j+n}+e_{j,i+n}+e_{j+n,i}+e_{i+n,j})\nonumber\\
&&-\frac{1}{\sqrt{2}}(e_{i,j+n}+e_{j,i+n}-e_{j+n,i}-e_{i+n,j})\otimes\frac{1}{\sqrt{2}}(e_{i,j+n}+e_{j,i+n}-e_{j+n,i}-e_{i+n,j})\nonumber\\
&&=e_{i,j+n}\otimes e_{j+n,i}+e_{i,j+n}\otimes e_{i+n,j}+e_{j,i+n}\otimes e_{j+n,i}+e_{j,i+n}\otimes e_{i+n,j}\nonumber\\
&&+e_{j+n,i}\otimes e_{i,j+n}+e_{j+n,i}\otimes e_{j,i+n}+e_{i+n,j}\otimes e_{i,j+n}+e_{i+n,j}\otimes e_{j,i+n}.
\end{eqnarray}

We also compute for $1\leq k\leq n$,
\begin{eqnarray}\label{prop-spn-proof-second}
&(e_{k, n+k}+e_{n+k, k})\otimes(e_{k, n+k}+e_{n+k, k})-(e_{k, n+k}-e_{n+k, k})\otimes(e_{k, n+k}-e_{n+k, k})\nonumber\\
&=2(e_{k,n+k}\otimes e_{n+k,k}+e_{n+k,k}\otimes e_{k,n+k}).
\end{eqnarray}

Again, considering another set of $n^{2}-n$ generators and applying (\ref{helpful-frml}), one obtains for $1\leq i<j\leq n$,
\begin{eqnarray}\label{prop-spn-proof-third}
&&\frac{1}{\sqrt{2}}(e_{ij}+e_{ji}-e_{i+n,j+n}-e_{j+n,i+n})\otimes\frac{1}{\sqrt{2}}(e_{ij}+e_{ji}-e_{i+n,j+n}-e_{j+n,i+n})\nonumber\\
&&-\frac{1}{\sqrt{2}}(e_{ij}-e_{ji}+e_{i+n,j+n}-e_{j+n,i+n})\otimes\frac{1}{\sqrt{2}}(e_{ij}-e_{ji}+e_{i+n,j+n}-e_{j+n,i+n})\nonumber\\
&&=e_{ij}\otimes e_{ji}-e_{ij}\otimes e_{i+n,j+n}-e_{j+n,i+n}\otimes e_{ji}+e_{j+n,i+n}\otimes e_{i+n,j+n}\nonumber\\
&&+e_{ji}\otimes e_{ij}-e_{ji}\otimes e_{j+n,i+n}-e_{i+n,j+n}\otimes e_{ij}+e_{i+n,j+n}\otimes e_{j+n,i+n}.
\end{eqnarray}

Finally, for $n$ diagonal generators of $Sp(2n,\mathbb{R})$, we obtain with $1\leq k\leq n$,
\begin{eqnarray}\label{prop-spn-proof-fourth}
\lefteqn{(e_{kk}-e_{k+n, k+n})\otimes(e_{kk}-e_{k+n, k+n})}\nonumber\\
&&=e_{kk}\otimes e_{kk}-e_{kk}\otimes e_{k+n,k+n}-e_{k+n,k+n}\otimes e_{kk}+e_{k+n,k+n}\otimes e_{k+n,k+n}.
\end{eqnarray}

Adding (\ref{prop-spn-proof-first}) to (\ref{prop-spn-proof-third}) and (\ref{prop-spn-proof-second}) to (\ref{prop-spn-proof-fourth}) followed by summing over $1\leq i<j\leq n$ and $1\leq k\leq n$, respectively and finally adding up the two summands, we obtain,

\begin{align}\label{final-exprssn-casimir-tensor-sp2n}
\Gamma &=\left[\sum_{1\leq i<j\leq n}(e_{i,j+n}\otimes e_{j+n,i}+e_{j,i+n}\otimes e_{i+n,j}+e_{j+n,i}\otimes e_{i,j+n}+e_{i+n,j}\otimes e_{j,i+n}\right.\nonumber\\
&+\left.e_{ij}\otimes e_{ji}+e_{j+n,i+n}\otimes e_{i+n,j+n}+e_{ji}\otimes e_{ij}+e_{i+n,j+n}\otimes e_{j+n,i+n})\right.\nonumber\\
&+\left.\sum_{1\leq k\leq n}(e_{k,n+k}\otimes e_{n+k,k}+e_{n+k,k}\otimes e_{k,n+k}+e_{kk}\otimes e_{kk}+e_{k+n,k+n}\otimes e_{k+n,k+n})\right]\nonumber\\
&+\left[\sum_{1\leq i<j\leq n}(e_{i,j+n}\otimes e_{i+n,j}+e_{j,i+n}\otimes e_{j+n,i}+e_{j+n,i}\otimes e_{j,i+n}+e_{i+n,j}\otimes e_{i,j+n}\nonumber\right.\\
&-\left.e_{ij}\otimes e_{i+n,j+n}-e_{j+n,i+n}\otimes e_{ji}-e_{ji}\otimes e_{j+n,i+n}-e_{i+n,j+n}\otimes e_{ij})\right.\nonumber\\
&+\left.\sum_{1\leq k\leq n}(e_{k,n+k}\otimes e_{n+k,k}+e_{n+k,k}\otimes e_{k,n+k}-e_{kk}\otimes e_{k+n,k+n}-e_{k+n,k+n}\otimes e_{kk})\right]\nonumber\\
&=P+\chi.
\end{align}
\qed

We require the following lemma to prove the main result regarding the Poisson bracket for $Sp(2n,\mathbb{R})$ monodromy matrices.
\begin{lem}\label{lem-relat-spn}
For $A, B\in Sp(2n,\mathbb{R})$, $\chi$ being the defect matrix as in proposition \ref{sp2n-prop} and $P$ being the permutation matrix, we have the following identity
\begin{equation}\label{useful-identity}
\tr_{12}[(A\otimes B)\chi]=-\tr(AB^{-1}).
\end{equation}
\end{lem}
\prf{.}
Given the $2n\times 2n$ symplectic matrix $B$, its inverse is given by the following sets of equations:

For the matrix entries with $1\leq i<j\leq n$,
\begin{equation}\label{inverse-symplectic}
\begin{aligned}
&(B^{-1})_{ij}=B_{j+n,i+n}, &\quad& (B^{-1})_{ji}=B_{i+n,j+n}, &\quad& (B^{-1})_{i,j+n}=-B_{j,i+n}\\
&(B^{-1})_{j,i+n}=-B_{i,j+n}, &\quad& (B^{-1})_{n+i,j}=-B_{j+n,i}, &\quad& (B^{-1})_{j+n,i}=-B_{n+i,j}\\
&(B^{-1})_{i+n,j+n}= B_{ji}, &\quad& (B^{-1})_{j+n,i+n}=B_{ij}.
\end{aligned}
\end{equation}
Whereas, for the matrix entries with $1\leq k\leq n$, one obtains
\begin{equation}\label{inverse-symplectic-diag}
\begin{aligned}
&(B^{-1})_{kk}=B_{k+n,k+n}, &\quad& (B^{-1})_{k,n+k}=-B_{k,n+k}\\
&(B^{-1})_{n+k,k}=-B_{n+k,k}, &\quad& (B^{-1})_{k+n,k+n}=B_{kk}.
\end{aligned}
\end{equation}

Using the explicit expression of the defect matrix $\chi$ given in (\ref{defect-matrix}) and that of the symplectic matrix $B^{-1}$ in (\ref{inverse-symplectic}) and (\ref{inverse-symplectic-diag}), one obtains
\begin{align}
&\tr_{12}[(A\otimes B)\chi]\nonumber\\
&=\sum_{1\leq i<j\leq n}(A_{j+n,i}B_{j,i+n}+A_{i+n,j}B_{i,j+n}+A_{i,j+n}B_{i+n,j}+A_{j,i+n}B_{j+n,i})\nonumber\\
&+\sum_{1\leq k\leq n}(A_{n+k,k}B_{k,n+k}+A_{k,n+k}B_{n+k,k})-\sum_{1\leq i<j\leq n}(A_{ji}B_{j+n,i+n}+A_{i+n,j+n}B_{ij})\nonumber\\
&-\sum_{1\leq i<j\leq n}(A_{ij}B_{i+n,j+n}+A_{j+n,i+n}B_{ji})-\sum_{1\leq k\leq n}(A_{kk}B_{k+n,k+n}+A_{k+n,k+n}B_{kk})\nonumber\\
&=-\sum_{1\leq i<j\leq n}[A_{j+n,i}(B^{-1})_{i,j+n}+A_{i+n,j}(B^{-1})_{j,i+n}+A_{i,j+n}(B^{-1})_{j+n,i}\nonumber\\
&\qquad\quad+A_{j,i+n}(B^{-1})_{i+n,j}]\nonumber\\
&-\sum_{1\leq k\leq n}[A_{n+k,k}(B^{-1})_{k,n+k}+A_{k,n+k}(B^{-1})_{n+k,k}]-\sum_{1\leq i<j\leq n}[A_{ji}(B^{-1})_{ij}\nonumber\\
&\qquad\quad+A_{i+n,j+n}(B^{-1})_{j+n,i+n}]\nonumber\\
&-\sum_{1\leq i<j\leq n}[A_{ij}(B^{-1})_{ji}+A_{j+n,i+n}(B^{-1})_{i+n,j+n}]-\sum_{1\leq k\leq n}[A_{kk}(B^{-1})_{kk}\nonumber\\
&\qquad\quad+A_{k+n,k+n}(B^{-1})_{k+n,k+n}]\nonumber\\
&=-\tr(AB^{-1}).
\end{align}
\qed

We now prove the main theorem concerning the Poisson bracket between traces of $Sp(2n,\mathbb{R})$ monodromy matrices.
\begin{thm}\label{goldman-brack-main-spn}
The Poisson bracket between traces of $Sp(2n,\mathbb{R})$ monodromy matrices $M_{\gamma_{1}}$ and $M_{\gamma_{2}}$ is given by
\begin{equation}\label{goldman-bracket-eqn-spn-monrdm}
\{\tr M_{\gamma_{1}},\tr M_{\gamma_{2}}\}=\frac{1}{2}\left(\tr M_{\gamma_{1}\circ\gamma_{2}}-\tr M_{\gamma_{1}\circ\gamma_{2}^{-1}}\right),
\end{equation}
where $M_{\gamma_{1}\circ\gamma_{2}}$ is an $Sp(2n,\mathbb{R})$ monodromy matrix computed along the loop $\gamma_{1}\circ\gamma_{2}$ as shown in figure \ref{fig:figure} while the monodromy matrix $M_{\gamma_{1}\circ\gamma_{2}^{-1}}$ is computed along the other loop $\gamma_{1}\circ\gamma_{2}^{-1}$ as given by figure \ref{fig:figure}.
\end{thm}
\prf{.}
Plugging the Casimir tensor $\Gamma$ (see (\ref{casimir-sp2n})) back in (\ref{eqnGldmn-brckt}) and using the identity from lemma \ref{lem-relat-spn}, one obtains
\begin{align}\label{final-formula-spn}
&\{\tr M_{\gamma_{1}},\tr M_{\gamma_{2}}\}\nonumber\\
&=\frac{1}{2}\tr_{12}[(T(0,x_{2})\widetilde{M}_{\gamma_{1}}T(x_{1},0)\otimes T(0,y_{2})\widetilde{M}_{\gamma_{2}}T(y_{1},0))(P+\chi)]\nonumber\\
&=\frac{1}{2}\tr M_{\gamma_{1}\circ\gamma_{2}}+\frac{1}{2}\tr_{12}[(T(0,x_{2})\widetilde{M}_{\gamma_{1}}T(x_{1},0)\otimes T(0,y_{2})\widetilde{M}_{\gamma_{2}}T(y_{1},0))\chi]\nonumber\\
&=\frac{1}{2}\tr M_{\gamma_{1}\circ\gamma_{2}}-\frac{1}{2}\tr[T(0,x_{2})\widetilde{M}_{\gamma_{1}}T(x_{1},0)T(0,y_{1})\widetilde{M}_{\gamma_{2}}^{-1}T(y_{2},0)]\nonumber\\
&=\frac{1}{2}\left(\tr M_{\gamma_{1}\circ\gamma_{2}}-\tr M_{\gamma_{1}\circ\gamma_{2}^{-1}}\right).
\end{align}
\qed

We now proceed to compute the Poisson bracket between traces of $SO(n)$ monodromy matrices. Let $e_{ij}$ denote an $n\times n$ matrix with 1 in $(i,j)$ entry and $0$ elsewhere. There are $\frac{(n^{2}-n)}{2}$ basis elements for the corresponding Lie algebra $\mathfrak{so(n)}$ given by
\begin{equation}\label{son bracket}
t_{a}=e_{ij}-e_{ji}\;\;\hbox{with}\;\;1\leq i<j \leq n.
\end{equation}
It can immediately be seen that the index $a$ runs from 1 to $\frac{n(n-1)}{2}$. In this case, $f(a)$, appearing in (\ref{orthonormal-basis-con}) is $-1$ for all $a$. The Casimir tensor $\Gamma$ for the Lie algebra $\mathfrak{so(n)}$ now reads
\begin{eqnarray}\label{casimir-son}
\Gamma&=&\sum_{i<j}-[(e_{ij}-e_{ji})\otimes (e_{ij}-e_{ji})]\nonumber\\
&=&-\sum_{i\neq j}(e_{ij}\otimes e_{ij})+\sum_{i\neq j}(e_{ij}\otimes e_{ji})\nonumber\\
&=&\sum_{i,j=1}^{n}(e_{ij}\otimes e_{ji})-\sum_{i,j=1}^{n}(e_{ij}\otimes e_{ij})\nonumber\\
&=&P+\chi,
\end{eqnarray}
where $P$ is the so-called Permutation matrix and $\chi$, which we refer to as the defect matrix for the Lie algebra $\mathfrak{so(n)}$, is given by
\begin{equation}\label{defect son}
\chi=-\sum_{i,j=1}^{n}(e_{ij}\otimes e_{ij}).
\end{equation}
We state the following lemma before deriving the Poisson bracket between traces of $SO(n)$ monodromy matrices.
\begin{lem}
Let $A,B\in SO(n)$ and $\chi$ be as given in (\ref{defect son}). Then the following holds
\begin{equation}
\tr_{12}[(A\otimes B)\chi]=-\tr(AB^{-1}).
\end{equation}
\end{lem}
\prf{.}
For any two $SO(n)$ matrices $A$ and $B$, we note that
\begin{eqnarray}\label{proof-lem-son}
\tr_{12}[(A\otimes B)\chi]&=&-\tr_{12}[(A\otimes B)(\sum_{i,j=1}^{n}e_{ij}\otimes e_{ij})]\nonumber\\
&=&-\tr_{12}(\sum_{i,j=1}^{n}Ae_{ij}\otimes Be_{ij})\nonumber\\
&=&-\sum_{i,j=1}^{n}\tr_{12}(Ae_{ij}\otimes Be_{ij})\nonumber\\
&=&-\sum_{i,j=1}^{n}\tr(Ae_{ij})\tr(Be_{ij})\nonumber\\
&=&-\sum_{i,j=1}^{n}A_{ji}B_{ji}\nonumber\\
&=&-\sum_{i,j=1}^{n}A_{ji}(B^{T})_{ij}\nonumber\\
&=&-\sum_{i,j=1}^{n}A_{ji}(B^{-1})_{ij}\nonumber\\
&=&-\tr(AB^{-1}).
\end{eqnarray}
\qed

Using the expression (\ref{casimir-son}) for the Casimir tensor $\Gamma$ of the Lie algebra $\mathfrak{so(n)}$ in the general formula (\ref{eqnGldmn-brckt}) and repeating the same computations as in the proof of theorem (\ref{goldman-brack-main-spn}), one obtains the Poisson brackets between the traces of $SO(n)$ monodromy matrices. We state this main result for the case of rotation group $SO(n)$ by means of the following theorem.
\begin{thm}\label{goldman-bracket-son}
The Poisson bracket between traces of $SO(n)$ monodromy matrices $M_{\gamma_{1}}$ and $M_{\gamma_{2}}$ is given by
\begin{equation}\label{goldman-bracket-eqn-son-monrdm}
\{\tr M_{\gamma_{1}},\tr M_{\gamma_{2}}\}=\frac{1}{2}\left(\tr M_{\gamma_{1}\circ\gamma_{2}}-\tr M_{\gamma_{1}\circ\gamma_{2}^{-1}}\right),
\end{equation}
where $M_{\gamma_{1}\circ\gamma_{2}}$ and $M_{\gamma_{1}\circ\gamma_{2}^{-1}}$ are $SO(n)$ monodromy matrices computed along the loops $\gamma_{1}\circ\gamma_{2}$ and $\gamma_{1}\circ\gamma_{2}^{-1}$, respectively. These loops are obtained by an appropriate resolution of intersection points and are illustrated in figure \ref{fig:figure}.
\end{thm}

We note that (\ref{goldman-bracket-gln-sun}), (\ref{goldman-bracket-sln-sun}), (\ref{goldman-bracket-eqn-spn-monrdm}) and (\ref{goldman-bracket-eqn-son-monrdm}) coincide with Goldman's formulae in (\cite{goldmant}, page 266). Here, we computed the Poisson bracket between traces of the monodromy matrices for a single point of transversal intersection. The proof for many intersection points follow similarly.

\section{Poisson bracket between \texorpdfstring{$G_{2}$}{G2}-gauge invariant observables}\label{sec:G2}
This section is dedicated to the computation of the Poisson bracket between $G_{2}$-gauge invariant observables which has not been considered in the literature so far. The starting point here is to compute the Poisson bracket between traces of $G_2$-monodromy matrices corresponding to two transversally intersecting loops on the underlying Riemann surface $\Sigma$. 

The exceptional real Lie group $G_2$ is $14$-dimensional. Below is a list of the appropriately normalized (in view of (\ref{orthonormal-basis-con})) $14$ basis elements of the corresponding exceptional real simple Lie algebra $\mathfrak{g}_{2}$ as given in \cite{eulerangle}.
\begin{align}\label{basis-elements-g2}
C_{1}&=\frac{1}{\sqrt{2}}(-e_{47}-e_{56}+e_{65}+e_{74})\nonumber\\
C_{2}&=\frac{1}{\sqrt{2}}(e_{46}-e_{57}-e_{64}+e_{75})\nonumber\\
C_{3}&=\frac{1}{\sqrt{2}}(-e_{45}+e_{54}-e_{67}+e_{76})\nonumber\\
C_{4}&=\frac{1}{\sqrt{2}}(e_{27}+e_{36}-e_{63}-e_{72})\nonumber\\
C_{5}&=\frac{1}{\sqrt{2}}(-e_{26}+e_{37}+e_{62}-e_{73})\nonumber\\
C_{6}&=\frac{1}{\sqrt{2}}(e_{25}-e_{34}+e_{43}-e_{52})\nonumber\\
C_{7}&=\frac{1}{\sqrt{2}}(-e_{24}-e_{35}+e_{42}+e_{53})\nonumber\\
C_{8}&=\frac{1}{\sqrt{6}}(-2e_{23}+2e_{32}+e_{45}-e_{54}-e_{67}+e_{76})\\
C_{9}&=\frac{1}{\sqrt{6}}(-2e_{12}+2e_{21}+e_{47}-e_{56}+e_{65}-e_{74})\nonumber\\
C_{10}&=\frac{1}{\sqrt{6}}(-2e_{13}+2e_{31}-e_{46}-e_{57}+e_{64}+e_{75})\nonumber\\
C_{11}&=\frac{1}{\sqrt{6}}(-2e_{14}-e_{27}+e_{36}+2e_{41}-e_{63}+e_{72})\nonumber\\
C_{12}&=\frac{1}{\sqrt{6}}(-2e_{15}+e_{26}+e_{37}+2e_{51}-e_{62}-e_{73})\nonumber\\
C_{13}&=\frac{1}{\sqrt{6}}(-2e_{16}-e_{25}-e_{34}+e_{43}+e_{52}+2e_{61})\nonumber\\
C_{14}&=\frac{1}{\sqrt{6}}(-2e_{17}+e_{24}-e_{35}-e_{42}+e_{53}+2e_{71}).\nonumber
\end{align}

The Casimir tensor $\Gamma$ corresponding to the fundamental representation of the Lie algebra $\mathfrak{g}_{2}$ is provided by the following proposition.
\begin{prop}\label{prop-casimir-g2}
The Casimir tensor $\Gamma$ for the fundamental representation of $\mathfrak{g}_{2}$ reads
\begin{equation}\label{casimir-g2}
\Gamma=\sum_{i,j=1}^{7}e_{ij}\otimes e_{ji}-\sum_{i,j=1}^{7}e_{ij}\otimes e_{ij}+\frac{1}{3}\sum_{i=1}^{7}\mathbb{O}_{i}\otimes\mathbb{O}_{i},
\end{equation}
where the matrices $\{\mathbb{O}_{i}\}$ are given by
\begin{eqnarray}\label{imaginary-octonions}
\begin{aligned}
\mathbb{O}_{1}&=&e_{32}-e_{23}+e_{54}-e_{45}-e_{76}+e_{67}\\
\mathbb{O}_{2}&=&e_{13}-e_{31}+e_{64}-e_{46}+e_{75}-e_{57}\\
\mathbb{O}_{3}&=&e_{21}-e_{12}+e_{74}-e_{47}-e_{65}+e_{56}\\
\mathbb{O}_{4}&=&e_{15}+e_{26}+e_{37}-e_{51}-e_{62}-e_{73}\\
\mathbb{O}_{5}&=&e_{41}-e_{72}+e_{63}-e_{14}-e_{36}+e_{27}\\
\mathbb{O}_{6}&=&e_{71}+e_{42}-e_{53}-e_{24}+e_{35}-e_{17}\\
\mathbb{O}_{7}&=&e_{52}-e_{61}+e_{43}-e_{34}-e_{25}+e_{16}.
\end{aligned}
\end{eqnarray}
\end{prop}
The proof is given in the Appendix \ref{sec-app}.

The skew-symmetric $7\times 7$ matrices $\{\mathbb{O}_{i}\}$ are reminiscent of the imaginary units of the normed division algebra of octonions, the multiplication table of which can be constructed out of the following relations (see \cite{octonionic-rep}):
\begin{equation}\label{octo-multplctn}
e_{i}e_{j}=-\delta_{ij}+\epsilon_{ijk}e_{k}\hspace{1in}(i,j,k=1,\dots,7),
\end{equation}
where $\epsilon_{ijk}$ is totally antisymmetric and is unity for the following set of combinations:
$$\{123, 145, 176, 246, 257, 347, 365 \}.$$ Note that $\{\mathbb{O}_{i}\}$ can be obtained from the $8\times 8$ matrix representations of $\{e_{i}\}$ (see \cite{octonionic-rep}) by deleting the first row and the first column of the respective matrices.

Use of proposition \ref{prop-casimir-g2} in (\ref{eqnGldmn-brckt}) yields the Poisson bracket between traces of $G_{2}$ monodromy matrices computed along homotopically inequivalent loops intersecting transversally at a ponit on $\Sigma$.The Poisson bracket between such canonical gauge invariant observables is provided by the following proposition.
\begin{prop}\label{Goldman-bracket-g2}
The Poisson bracket between traces of $G_{2}$ monodromy matrices $M_{\gamma_{1}}$ and $M_{\gamma_{2}}$, computed along homotopically inequivalent and transversally intersecting loops $\gamma_{1}$ and $\gamma_{2}$, respectively, is given by
\begin{equation}\label{goldman-bracket-eqn-g2-monrdm}
\{\tr M_{\gamma_{1}},\tr M_{\gamma_{2}}\}=\frac{1}{2}\left[\tr M_{\gamma_{1}\circ\gamma_{2}}-\tr M_{\gamma_{1}\circ\gamma_{2}^{-1}}+\frac{1}{3}\sum_{i=1}^{7}\tr(M_{\gamma_{1}}\mathbb{O}_{i})\tr(M_{\gamma_{2}}\mathbb{O}_{i})\right],
\end{equation}
where $M_{\gamma_{1}\circ\gamma_{2}}$ and $M_{\gamma_{1}\circ\gamma_{2}^{-1}}$ are $G_{2}$ monodromy matrices computed along the loop $\gamma_{1}\circ\gamma_{2}$ and $\gamma_{1}\circ\gamma_{2}^{-1}$, respectively. These loops are illustrated in figure \ref{fig:figure}. The skew-symmetric matrices $\{\mathbb{O}_{i}\}$ are as given in (\ref{imaginary-octonions}).
\end{prop}
\prf.
\begin{eqnarray}\label{proof-g2-goldman-brckt}
\lefteqn{\{\tr M_{\gamma_{1}},\tr M_{\gamma_{2}}\}}\nonumber\\
&&=\frac{1}{2}\tr_{12}[(T(0,x_{2})\widetilde{M}_{\gamma_{1}}T(x_{1},0)\otimes T(0,y_{2})\widetilde{M}_{\gamma_{2}}T(y_{1},0))\Gamma]\nonumber\\
&&=\frac{1}{2}\tr_{12}\left[(T(0,x_{2})\widetilde{M}_{\gamma_{1}}T(x_{1},0)\otimes T(0,y_{2})\widetilde{M}_{\gamma_{2}}T(y_{1},0 ))\right.\nonumber\\
&&\;\;\;\;\left.\times\left(\sum_{i,j=1}^{7}e_{ij}\otimes e_{ji}-\sum_{i,j=1}^{7}e_{ij}\otimes e_{ij}+\frac{1}{3}\sum_{i=1}^{7}\mathbb{O}_{i}\otimes\mathbb{O}_{i}\right)\right]\nonumber\\
&&=\frac{1}{2}\tr_{12}[(T(0,x_{2})\widetilde{M}_{\gamma_{1}}T(x_{1},0)\otimes T(0,y_{2})\widetilde{M}_{\gamma_{2}}T(y_{1},0 ))P]\nonumber\\
&&\;\;-\frac{1}{2}\tr_{12}\left[\sum_{i,j=1}^{7}(T(0,x_{2})\widetilde{M}_{\gamma_{1}}T(x_{1},0)e_{ij}\otimes T(0,y_{2})\widetilde{M}_{\gamma_{2}}T(y_{1},0 )e_{ij})\right]\nonumber\\
&&\;\;+\frac{1}{6}\tr_{12}\left[\sum_{i=1}^{7}(T(0,x_{2})\widetilde{M}_{\gamma_{1}}T(x_{1},0)\mathbb{O}_{i}\otimes  T(0,y_{2})\widetilde{M}_{\gamma_{2}}T(y_{1},0 )\mathbb{O}_{i})\right]\nonumber\\
&&=\frac{1}{2}\tr M_{\gamma_{1}\circ\gamma_{2}}-\frac{1}{2}\sum_{i,j=1}^{7}(T(0,x_{2})\widetilde{M}_{\gamma_{1}}T(x_{1},0))_{ji}(T(0,y_{2})\widetilde{M}_{\gamma_{2}}T(y_{1},0))_{ji}\nonumber\\
&&\;\;+\frac{1}{6}\sum_{i=1}^{7}\tr(M_{\gamma_{1}}\mathbb{O}_{i})\tr(M_{\gamma_{2}}\mathbb{O}_{i})\nonumber\\
&&=\frac{1}{2}\tr M_{\gamma_{1}\circ\gamma_{2}}-\frac{1}{2}\sum_{i,j=1}^{7}(T(0,x_{2})\widetilde{M}_{\gamma_{1}}T(x_{1},0))_{ji}(T(0,y_{1})\widetilde{M}_{\gamma_{2}}^{-1}T(y_{2},0))_{ij}\nonumber\\
&&\;\;+\frac{1}{6}\sum_{i=1}^{7}\tr(M_{\gamma_{1}}\mathbb{O}_{i})\tr(M_{\gamma_{2}}\mathbb{O}_{i})\nonumber\\
&&=\frac{1}{2}\tr M_{\gamma_{1}\circ\gamma_{2}}-\frac{1}{2}\tr(T(0,x_{2})\widetilde{M}_{\gamma_{1}}T(x_{1},0)T(0,y_{1})\widetilde{M}_{\gamma_{2}}^{-1}T(y_{2},0))\nonumber\\
&&\;\;+\frac{1}{6}\sum_{i=1}^{7}\tr(M_{\gamma_{1}}\mathbb{O}_{i})\tr(M_{\gamma_{2}}\mathbb{O}_{i})\nonumber\\
&&=\frac{1}{2}\left[\tr M_{\gamma_{1}\circ\gamma_{2}}-\tr M_{\gamma_{1}\circ\gamma_{2}^{-1}}+\frac{1}{3}\sum_{i=1}^{7}\tr(M_{\gamma_{1}}\mathbb{O}_{i})\tr(M_{\gamma_{2}}\mathbb{O}_{i})\right].
\end{eqnarray}
\qed
\begin{remark}
We remark here that the term $\sum\limits_{i=1}^{7}\tr(M_{\gamma_{1}}\mathbb{O}_{i})\tr(M_{\gamma_{2}}\mathbb{O}_{i})$  in (\ref{goldman-bracket-eqn-g2-monrdm}) is $G_{2}$-gauge invariant while none of the terms $\{\tr(M_{\gamma_{1}}\mathbb{O}_{i})\}$ is, as can easily be verified with the help of different 1-parametric subgroups of $G_{2}$.
\end{remark}

We can obtain an infinite set of such gauge invariant observables using the following Lemma (\cite{mathoverflow}):
\begin{lem}[Due to Jos\'{e} Figueroa-O'Farrill]
\label{lemma-transfrng-left-to-right-act}
Let $g$ be an element of $G_2$ in its $7$-dimensional fundamental representation and denote the relevant matrix by $g_{ij}$. Given the $7$ skew-symmetric $7\times 7$ matrices $\mathbb{O}_{i}$'s as enumerated in (\ref{imaginary-octonions}), the following holds:
\begin{equation}\label{lemma-transfr-action}
g\mathbb{O}_{i}g^{-1}=\sum_{k=1}^{7}\mathbb{O}_{k}g_{ki}.
\end{equation}
\end{lem}

\prf{.}
The proof is based on the fact that $G_2$ is the automorphism group of octonions.

Let $e_{1},e_{2},..,e_{7}$ denote the $7$ imaginary octonions that obey the multiplication rule given by (\ref{octo-multplctn}) and 1 denote the identity. Using (\ref{imaginary-octonions}) together with (\ref{octo-multplctn}), one obtains
\begin{equation}\label{useful-formula}
e_{i}e_{j}=\sum_{k=1}^{7}e_{k}(\mathbb{O}_{i})_{kj}-\delta_{ij}1.
\end{equation}
Now the action of $g\in G_{2}$ on an octonionic imaginary unit by means of its matrix representation $g_{ij}$ reads
\begin{equation}\label{change-of-basis}
g(e_i)=\sum_{j=1}^{7}e_{j}g_{ji}.
\end{equation}
Let us now apply $g\in G_{2}$ on both sides of (\ref{useful-formula}). The left hand side then gives
\begin{eqnarray}\label{left-side}
g(e_{i}e_{j})&=&g(e_i)g(e_j)\nonumber\\
&=&\sum_{k,l=1}^{7}(e_{k}e_{l})g_{ki}g_{lj}\nonumber\\
&=&\sum_{k,l=1}^{7}\left(\sum_{m=1}^{7}e_{m}(\mathbb{O}_{k})_{ml}-\delta_{kl}1\right)g_{ki}g_{lj}\nonumber\\
&=&\sum_{k,m=1}^{7}e_{m}(\mathbb{O}_{k}g)_{mj}g_{ki}-\delta_{ij}1,
\end{eqnarray}
where the orthogonality of $g\in G_{2}$ is exploited. Similarly, application of $g$ on the right hand side of (\ref{useful-formula}) yields
\begin{eqnarray}\label{right-side}
\sum_{k=1}^{7}g(e_{k}(\mathbb{O}_{i})_{kj}-\delta_{ij}g(1))
&=&\sum_{k,m=1}^{7}e_{m}g_{mk}(\mathbb{O}_{i})_{kj}-\delta_{ij}1\nonumber\\
&=&\sum_{m=1}^{7}e_{m}(g{\mathbb{O}}_{i})_{mj}-{\delta}_{ij}1,
\end{eqnarray}
where we have used the fact that $g(1)=1$. Now comparing (\ref{left-side}) with (\ref{right-side}), one finds
\begin{equation}\label{almost-final-expression}
\sum_{k=1}^{7}{({\mathbb{O}}_{k}g)}_{mj}{g}_{ki}=(g{\mathbb{O}}_{i})_{mj}.
\end{equation}
Multiplying both sides of (\ref{almost-final-expression}) by $g^{-1}_{jl}$ and summing over $j$, we obtain the following for any $m$ and $l$:
\begin{equation}\label{final-exp}
(\sum_{k=1}^{7}{\mathbb{O}}_{k}{g}_{ki})_{ml}=(g{\mathbb{O}}_{i}g^{-1})_{ml}.
\end{equation}
\qed.

Now we proceed to state the main result of the paper by means of a theorem that concerns an infinite set of $G_2$-gauge invariant observables. The theorem is as follows
\begin{Theo}\label{G2-gauge-inv-obsrvbl}
Given 4 non-negative integers $r$, $s$, $n_1$ and $n_2$ satisfying $r\leq n_{1}$, $s\leq n_{2}$ and two $(0,1)$-matrices $K$ and $Q$ of order $t\times n_{1}$ and $t\times(2n_{2}-s)$, respectively and $t$ being another positive integer satisfying $t\leq n_{1}+2n_{2}$, the following expression involving monodromy matrices $M_{\gamma_{j}}$ corresponding to pairwise transversally intersecting $n_{1}+t$ loops $\gamma_{j}$'s with $j=1,2,...,n_{1}+t$, is a $G_2$ gauge invariant observable
\begin{align}\label{G2-gauge-invrnt-obsrvbl-exprssn}
&\sum\limits_{l_{j}\in\{1,2,...,7\}}\tr (M_{\gamma_1}\mathbb{O}_{l_1})\tr(M_{\gamma_2}\mathbb{O}_{l_2})...\tr(M_{\gamma_r}\mathbb{O}_{l_r})\tr(M_{\gamma_{r+1}}\mathbb{O}_{l_{r+1}})...\tr(M_{\gamma_{n_{1}}}\mathbb{O}_{l_{n_{1}}})\nonumber\\
&\times\tr\left(M_{\gamma_{n_{1}+1}}\mathbb{O}^{K_{1,1}}_{l_{1}}...\mathbb{O}^{K_{1,r}}_{l_{r}}\mathbb{O}^{K_{1,r+1}}_{l_{n_{1}+1}}...\mathbb{O}^{K_{1,n_{1}}}_{l_{2n_{1}-r}}\mathbb{O}^{Q_{1,1}}_{l_{2n_{1}-r+1}}...\mathbb{O}^{Q_{1,s}}_{l_{2n_{1}-r+s}}\mathbb{O}^{Q_{1,s+1}}_{l_{2n_{1}-r+s+1}}\mathbb{O}^{Q_{1,s+2}}_{l_{2n_{1}-r+n_{2}+1}}\right.\nonumber\\
&
\left.\hspace{3.4in}...\mathbb{O}^{Q_{1,s+(2n_{2}-2s-1)}}_{l_{2n_{1}-r+s+(n_{2}-s)}}\mathbb{O}^{Q_{1,s+(2n_{2}-2s)}}_{l_{2n_{1}-r+n_{2}+(n_{2}-s)}}\right)\nonumber\\
&\times\tr\left(M_{\gamma_{n_{1}+2}}\mathbb{O}^{K_{2,1}}_{l_{1}}...\mathbb{O}^{K_{2,r}}_{l_{r}}\mathbb{O}^{K_{2,r+1}}_{l_{n_{1}+1}}...\mathbb{O}^{K_{2,n_{1}}}_{l_{2n_{1}-r}}\mathbb{O}^{Q_{2,1}}_{l_{2n_{1}-r+1}}...\mathbb{O}^{Q_{2,s}}_{l_{2n_{1}-r+s}}\mathbb{O}^{Q_{2,s+1}}_{l_{2n_{1}-r+s+1}}\mathbb{O}^{Q_{2,s+2}}_{l_{2n_{1}-r+n_{2}+1}}\right.\nonumber\\
&\left.\hspace{3.4in}...\mathbb{O}^{Q_{2,s+(2n_{2}-2s-1)}}_{l_{2n_{1}-r+s+(n_{2}-s)}}\mathbb{O}^{Q_{2,s+(2n_{2}-2s)}}_{l_{2n_{1}-r+n_{2}+(n_{2}-s)}}\right)\nonumber\\
&\vdots\hspace{1in}\vdots\hspace{1in}\vdots\hspace{1in}\vdots\hspace{1in}\vdots\hspace{1in}\vdots\nonumber\\
&\vdots\hspace{1in}\vdots\hspace{1in}\vdots\hspace{1in}\vdots\hspace{1in}\vdots\hspace{1in}\vdots\nonumber\\
&\times\tr\left(M_{\gamma_{n_{1}+t}}\mathbb{O}^{K_{t,1}}_{l_{1}}...\mathbb{O}^{K_{t,r}}_{l_{r}}\mathbb{O}^{K_{t,r+1}}_{l_{n_{1}+1}}...\mathbb{O}^{K_{t,n_{1}}}_{l_{2n_{1}-r}}\mathbb{O}^{Q_{t,1}}_{l_{2n_{1}-r+1}}...\mathbb{O}^{Q_{t,s}}_{l_{2n_{1}-r+s}}\mathbb{O}^{Q_{t,s+1}}_{l_{2n_{1}-r+s+1}}\mathbb{O}^{Q_{2,s+2}}_{l_{2n_{1}-r+n_{2}+1}}\right.\nonumber\\
&\left.\hspace{3.4in}...\mathbb{O}^{Q_{t,s+(2n_{2}-2s-1)}}_{l_{2n_{1}-r+s+(n_{2}-s)}}\mathbb{O}^{Q_{t,s+(2n_{2}-2s)}}_{l_{2n_{1}-r+n_{2}+(n_{2}-s)}}\right)\nonumber\\
&\times (\alpha^{1})_{l_{r+1},l_{n_{1}+1}}(\alpha^{2})_{l_{r+2},l_{n_{1}+2}}...(\alpha^{n_{1}-r})_{l_{n_{1}},l_{2n_{1}-r}}\nonumber\\
&\times (\beta^{1})_{l_{2n_{1}-r+s+1},l_{2n_{1}-r+n_{2}+1}}(\beta^{2})_{l_{2n_{1}-r+s+2},l_{2n_{1}-r+n_{2}+2}}...(\beta^{n_{2}-s})_{l_{2n_{1}-r+n_{2}},l_{2n_{1}+2n_{2}-r-s}},
\end{align}
where $\alpha^{1}$, $\alpha^{2}$,...,$\alpha^{n_{1}-r}$ and $\beta^{1}$, $\beta^{2}$,...,$\beta^{n_{2}-s}$ are in 7-dimensional fundamental representation of $G_{2}$. The noncommuting indeterminates $\mathbb{O}_{l_{j}}$'s take their values in $\{\mathbb{O}_{1},\mathbb{O}_{2},...,\mathbb{O}_{7}\}$ (see \ref{imaginary-octonions}). Also, here, each column of the $(0,1)$-matrix $K$ has exactly one entry equal to 1. The $(0,1)$-matrix $Q$ of order $t\times(2n_{2}-s)$, on the other hand, has exactly two entries equal to 1 in each column of the $t\times s$-block and one entry equal to 1 in each column of the adjacent $t\times (2n_{2}-2s)$-block. 
\end{Theo}

The proof is given in the Appendix \ref{sec-app}.

\begin{remark}
It is worth remarking at this point that for a fixed tuple of non-negative integers $(r,n_1,s,n_2,t)$, with $r\leq n_{1}$, $s\leq n_{2}$ and $t\leq n_{1}+2n_{2}$, there are only finitely many choices for the $(0,1)$-matrices $K$ and $Q$ to obtain various $G_2$-gauge invariant observables. By varying $r$, $n_1$, $s$, $n_2$ and $t$ subject to the above mentioned constraints, one thus obtains an infinite set of such gauge invariant observables.
\end{remark}

We will refer to the observables obtained in Theorem \ref{G2-gauge-inv-obsrvbl} as {\em exotic gauge invariant observables}. 
The Poisson bracket between two such exotic $G_2$ gauge invariant observables can be seen to be a linear combination of exotic $G_2$ gauge invariant observables by using formula (\ref{finlmondrmcomptn}) and Lemma \ref{lemma-transfrng-left-to-right-act} repeatedly. On the other hand, the Poisson bracket between a canonical $G_2$ gauge invariant observable (trace of a $G_2$ monodromy matrix) and an exotic $G_2$ gauge invariant observable can again be shown to be a linear combination of exotic $G_2$ gauge invariant observables. We will only prove the latter fact for the fact that the Poisson bracket between two exotic $G_2$-gauge invariant observables being a linear combination of gauge invariant observables of the same type can be proven using exactly similar techniques.

Let us fix a few notations first. Denote the trace monomial in theorem (\ref{G2-gauge-inv-obsrvbl}) by $F(r,n_1,s,n_2,t;K(r,n_1,t),Q(s,n_2,t);\gamma_1,\gamma_2,...,\gamma_{n_{1}+t};l_{j})$. Deletion of a trace term involving monodromy around the loop $\gamma_i$ in the expression of the sum over trace monomials (see \ref{G2-gauge-invrnt-obsrvbl-exprssn}) will be denoted by $F(r,n_1,s,n_2,t;K(r,n_1,t),Q(s,n_2,t);\gamma_1,...,\hat{\gamma}_i,...,\gamma_{n_{1}+t};l_j)$. In what follows, we will adopt the shorthand notation $F(\gamma_1,\gamma_2,...,\gamma_{n_{1}+t};l_j)$ where no confusion with $r$, $n_1$, $s$, $n_2$, $t$, $K(r,n_1,t)$ and $Q(s,n_2,t)$ arises, i.e., they remain unchanged. One therefore obtains
\begin{align}\label{proof-closedness}
&\{\tr M_{\gamma},\sum_{l_{j}\in\{1,2,...,7\}}F(r,n_1,s,n_2,t;K(r,n_1,t),Q(s,n_2,t);\gamma_1,\gamma_2,...,\gamma_{n_{1}+t};l_{j})\}\nonumber\\
&=\sum_{l_{j}\in\{1,2,...,7\}}\{\tr M_{\gamma},\tr(M_{\gamma_1}\mathbb{O}_{l_1})\}F(\hat{\gamma}_{1},\gamma_2,...,\gamma_{n_{1}+t};l_j)+\cdots\nonumber\\
&+\sum_{l_{j}\in\{1,2,...,7\}}\{\tr M_{\gamma},\tr(M_{\gamma_{r+1}}\mathbb{O}_{l_{r+1}})\}F(\gamma_1,\cdots,\hat{\gamma}_{r+1},\cdots,\gamma_{n_{1}+t};l_j)+\cdots\nonumber\\
&+\sum_{l_{j}\in\{1,2,...,7\}}\{\tr M_{\gamma},\tr(M_{\gamma_{n_{1}+1}}\mathbb{O}^{K_{1,1}}_{l_1}\cdots\mathbb{O}^{K_{1,n_1}}_{l_{2n_{1}-r}}\mathbb{O}^{Q_{1,1}}_{l_{2n_{1}-r+1}}\cdots\mathbb{O}^{Q_{1,2n_{2}-s}}_{l_{2n_{1}+2n_{2}-r-s}})\}\nonumber\\
&\times F(\gamma_1,\cdots,\hat{\gamma}_{n_{1}+1},\cdots,\gamma_{n_{1}+t};l_j)+\cdots\nonumber\\
&+\sum_{l_{j}\in\{1,2,...,7\}}\{\tr M_{\gamma},\tr(M_{\gamma_{n_{1}+t}}\mathbb{O}^{K_{t,1}}_{l_1}\cdots\mathbb{O}^{K_{t,n_1}}_{l_{2n_{1}-r}}\mathbb{O}^{Q_{t,1}}_{l_{2n_{1}-r+1}}\cdots\mathbb{O}^{Q_{t,2n_{2}-s}}_{l_{2n_{1}+2n_{2}-r-s}})\}\nonumber\\
&\times F(\gamma_1,\gamma_2,\cdots,\hat{\gamma}_{n_{1}+t};l_j)
\end{align}

The claim now is that each of the $(n_{1}+t)$ summands above is a linear combination of exotic $G_2$-gauge invariant expressions of the type given by (\ref{G2-gauge-invrnt-obsrvbl-exprssn}). It is,  therefore, sufficient to prove the claim for the following three summands in (\ref{proof-closedness}):
\begin{itemize}
\item $\sum\limits_{l_{j}\in\{1,2,...,7\}}\{\tr M_{\gamma},\tr(M_{\gamma_1}\mathbb{O}_{l_1})\}F(\hat{\gamma}_{1},\gamma_2,...,\gamma_{n_{1}+t};l_j)$, 
\item $\sum\limits_{l_{j}\in\{1,2,...,7\}}\{\tr M_{\gamma},\tr(M_{\gamma_{r+1}}\mathbb{O}_{l_{r+1}})\}F(\gamma_{1},\cdots,\hat{\gamma}_{r+1},\cdots,\gamma_{n_{1}+t};l_j)$,
\item $\sum\limits_{l_{j}\in\{1,2,...,7\}}\{\tr M_{\gamma},\tr(M_{\gamma_{n_{1}+t}}\mathbb{O}^{K_{t,1}}_{l_1}\cdots\mathbb{O}^{K_{t,n_1}}_{l_{2n_{1}-r}}\mathbb{O}^{Q_{t,1}}_{l_{2n_{1}-r+1}}\cdots\mathbb{O}^{Q_{t,2n_{2}-s}}_{l_{2n_{1}+2n_{2}-r-s}})\}\\
\times F(\gamma_1,\gamma_2,\cdots,\hat{\gamma}_{n_{1}+t};l_j)$,
\end{itemize}

Now using formula (\ref{finlmondrmcomptn}) and lemma \ref{lemma-transfrng-left-to-right-act}, one obtains
\begin{align}\label{first-summand}
&\sum_{l_{j}}\{\tr M_{\gamma},\tr(M_{\gamma_1}\mathbb{O}_{l_1})\}F(r,n_{1},s,n_{2},t;K(r,n_{1},t),Q(s,n_{2},t);\hat{\gamma}_{1},\gamma_2,\cdots,\gamma_{n_{1}+t};l_j)\nonumber\\
&=\frac{1}{2}\sum_{l_{j}}F(r,n_{1},s,n_{2},t;K(r,n_{1},t),Q(s,n_{2},t);\gamma\circ\gamma_{1},\gamma_2,\cdots,\gamma_{n_{1}+t};l_j)\nonumber\\
&+\frac{1}{2}\sum_{l_{j}}F(r,n_{1},s,n_{2},t;K(r,n_{1},t),Q(s,n_{2},t);\gamma\circ\gamma_{1}^{-1},\gamma_2,\cdots,\gamma_{n_{1}+t};l_j)\nonumber\\
&+\frac{1}{6}\sum_{l_{j}}F(r-1,n_{1},s+1,n_{2}+1,t+1;K^{\prime}(r-1,n_{1},t+1),Q^{\prime}(s+1,n_{2}+1,t+1)\nonumber\\
&\hspace{.2in};\underbrace{\gamma_{2},\cdots,\gamma_{r}}_{r-1},\underbrace{\gamma,\gamma_{r+1},\cdots,\gamma_{n_1}}_{n_{1}-r+1},\underbrace{\gamma_{1},\gamma_{n_{1}+1},\cdots,\gamma_{n_{1}+t}}_{t+1};l_j),
\end{align}
where $K^{\prime}(r-1,n_{1},t+1)$ and $Q^{\prime}(s+1,n_{2}+1,t+1)$ are $(0,1)$-matrices of order $(t+1)\times n_{1}$ and $(t+1)\times (2n_{2}-s+1)$, respectively, as in theorem \ref{G2-gauge-inv-obsrvbl}.
By using formula (\ref{finlmondrmcomptn}) together with lemma \ref{lemma-transfrng-left-to-right-act}, one also obtains the following:
\begin{align}\label{second-summand}
&\sum_{l_{j}}\{\tr M_{\gamma},\tr(M_{\gamma_{r+1}}\mathbb{O}_{l_{r+1}})\}F(r,n_{1},s,n_{2},t;K(r,n_{1},t),Q(s,n_{2},t)\nonumber\\
&\hspace{.2in};\gamma_{1},\cdots,\hat{\gamma}_{r+1},\cdots,\gamma_{n_{1}+t};l_j)\nonumber\\
&=\frac{1}{2}\sum_{l_{j}}F(r,n_{1},s,n_{2},t;K(r,n_{1},t),Q(s,n_{2},t);\gamma_{1},\cdots,\gamma_{r},\gamma\circ\gamma_{r+1},\gamma_{r+2},\cdots,\gamma_{n_{1}+t};l_j)\nonumber\\
&+\frac{1}{2}\sum_{l_{j}}F(r,n_{1},s,n_{2},t;K(r,n_{1},t),Q(s,n_{2},t);\gamma_{1},\cdots,\gamma_{r},\gamma\circ\gamma_{r+1}^{-1},\gamma_{r+2},\cdots,\gamma_{n_{1}+t};l_j)\nonumber\\
&+\frac{1}{6}\sum_{l_{j}}F(r,n_{1},s,n_{2}+1,t+1;K^{\prime\prime}(r,n_{1},t+1),Q^{\prime\prime}(s,n_{2}+1,t+1)\nonumber\\
&\hspace{.2in};\underbrace{\gamma_{1},\cdots,\gamma_{r}}_{r},\underbrace{\gamma,\gamma_{r+2},\cdots,\gamma_{n_1}}_{n_{1}-r},\underbrace{\gamma_{r+1},\gamma_{n_{1}+1},\cdots,\gamma_{n_{1}+t}}_{t+1};l_j),
\end{align}
where the $(0,1)$-matrices $K^{\prime\prime}(r,n_{1},t+1)$ and $Q^{\prime\prime}(s,n_{2}+1,t+1)$, that are of order $(t+1)\times n_{1}$ and $(t+1)\times (2n_{2}-s+2)$, respectively, satisfy the properties enumerated in theorem \ref{G2-gauge-inv-obsrvbl}. Finally, for the third summand, one obtains
\begin{align}\label{third-summand}
&\sum\limits_{l_{j}}\{\tr M_{\gamma},\tr(M_{\gamma_{n_{1}+t}}\mathbb{O}^{K_{t,1}}_{l_1}\cdots\mathbb{O}^{K_{t,n_1}}_{l_{2n_{1}-r}}\mathbb{O}^{Q_{t,1}}_{l_{2n_{1}-r+1}}\cdots\mathbb{O}^{Q_{t,2n_{2}-s}}_{l_{2n_{1}+2n_{2}-r-s}})\}\nonumber\\
&\times F(r,n_{1},s,n_{2},t;K(r,n_{1},t),Q(s,n_{2},t);\gamma_1,\gamma_2,\cdots,\hat{\gamma}_{n_{1}+t};l_j)\nonumber\\
&=\frac{1}{2}\sum_{l_{j}}F(r,n_{1},s,n_{2},t;K(r,n_{1},t),Q(s,n_{2},t);\gamma_{1},\cdots,\gamma_{n_{1}},\cdots,\gamma_{n_{1}+t-1},\gamma\circ\gamma_{n_{1}+t};l_j)\nonumber\\
&+\frac{1}{2}\sum_{l_{j}}F(r,n_{1},s,n_{2},t;K(r,n_{1},t),Q(s,n_{2},t);\gamma_{1},\cdots,\gamma_{n_{1}},\cdots,\gamma_{n_{1}+t-1},\gamma\circ\gamma_{n_{1}+t}^{-1};l_j)\nonumber\\
&+\frac{1}{6}\sum_{l_{j}}F(r,n_{1}+1,s,n_{2},t;K^{\prime\prime\prime}(r,n_{1}+1,t),Q^{\prime\prime\prime}(s,n_{2},t)\nonumber\\
&\hspace{.2in};\underbrace{\gamma_{1},\cdots,\gamma_{r}}_{r},\underbrace{\gamma,\gamma_{r+1},\cdots,\gamma_{n_1}}_{n_{1}-r+1},\underbrace{\gamma_{n_{1}+1},\cdots,\gamma_{n_{1}+t}}_{t};l_j),
\end{align}
where $K^{\prime\prime\prime}(r,n_{1}+1,t)$ and $Q^{\prime\prime\prime}(s,n_{2},t)$ are certain $(0,1)$-matrices of order $t\times(n_{1}+1)$ and $t\times(2n_{2}-s)$, respectively, satisfying the required properties as enumerated in theorem \ref{G2-gauge-inv-obsrvbl}. 

Let us, give a few concrete examples of such exotic gauge invariant observables and elucidate what roles the non-negative parameters $r$, $n_1$, $s$, $n_2$, $t$ and the $(0,1)$-matrices $K$ and $Q$, enumerated in theorem \ref{G2-gauge-inv-obsrvbl}, have got to play in these contexts.

\subsection{Examples of exotic gauge invariant observables}\label{subsec-exmpl-gauge-invrnt-ob}
We encountered the first instance of exotic $G_2$-gauge invariant observables in proposition \ref{Goldman-bracket-g2}, namely, the term $\sum\limits_{i=1}^{7}\tr(M_{\gamma_1}\mathbb{O}_{i})\tr(M_{\gamma_2}\mathbb{O}_{i})$ while computing the Poisson bracket between traces of $G_2$ monodromy matrices. In the preceding example, $r=n_{1}=1$, $s=n_{2}=0$ and $t=1$ so that we have the $(0,1)$-matrix  
$K$ to be just 1 as implicated by theorem \ref{G2-gauge-inv-obsrvbl}. 

The second example, to be considered here, emerges from the explicit computation of the following expression for pairwise transversally intersecting 4 loops $\{\gamma_i\}$ on $\Sigma$: $\{\sum\limits_{i=1}^{7}\tr(M_{\gamma_1}\mathbb{O}_{i})\tr(M_{\gamma_2}\mathbb{O}_{i}),\sum\limits_{j=1}^{7}\tr(M_{\gamma_3}\mathbb{O}_{j})\tr(M_{\gamma_4}\mathbb{O}_{j})\}$. Using formula (\ref{finlmondrmcomptn}) together with lemma \ref{lemma-transfrng-left-to-right-act}, one obtains
\begin{align}
&\{\sum\limits_{i=1}^{7}\tr(M_{\gamma_1}\mathbb{O}_{i})\tr(M_{\gamma_2}\mathbb{O}_{i}),\sum\limits_{j=1}^{7}\tr(M_{\gamma_3}\mathbb{O}_{j})\tr(M_{\gamma_4}\mathbb{O}_{j})\}\nonumber\\
&=\frac{1}{2}\sum_{i,j,k=1}^{7}\alpha_{kj}\tr(M_{\gamma_{1}\circ\gamma_{3}}\mathbb{O}_{i}\mathbb{O}_{k})\tr(M_{\gamma_{2}}\mathbb{O}_{i})\tr(M_{\gamma_{4}}\mathbb{O}_{j})\label{exmpl-gauge-invrnt-obsrvbl-1}\\
&+\frac{1}{2}\sum_{i,j,k=1}^{7}\beta_{kj}\tr(M_{\gamma_{1}\circ\gamma_{3}^{-1}}\mathbb{O}_{i}\mathbb{O}_{k})\tr(M_{\gamma_{2}}\mathbb{O}_{i})\tr(M_{\gamma_{4}}\mathbb{O}_{j})\label{exmpl-gauge-invrnt-obsrvbl-2}\\
&+\frac{1}{6}\sum_{i,j,m,l=1}^{7}\gamma_{ml}\tr(M_{\gamma_{1}}\mathbb{O}_{i}\mathbb{O}_{l})\tr(M_{\gamma_{3}}\mathbb{O}_{j}\mathbb{O}_{m})\tr(M_{\gamma_{2}}\mathbb{O}_{i})\tr(M_{\gamma_{4}}\mathbb{O}_{j})\label{exmpl-gauge-invrnt-obsrvbl-3}\\
&+\frac{1}{2}\sum_{i,j,k=1}^{7}\alpha^{\prime}_{kj}\tr(M_{\gamma_{1}\circ\gamma_{4}}\mathbb{O}_{i}\mathbb{O}_{k})\tr(M_{\gamma_{2}}\mathbb{O}_{i})\tr(M_{\gamma_{3}}\mathbb{O}_{j})\nonumber\\
&+\frac{1}{2}\sum_{i,j,k=1}^{7}\beta^{\prime}_{kj}\tr(M_{\gamma_{1}\circ\gamma_{4}^{-1}}\mathbb{O}_{i}\mathbb{O}_{k})\tr(M_{\gamma_{2}}\mathbb{O}_{i})\tr(M_{\gamma_{3}}\mathbb{O}_{j})\nonumber\\
&+\frac{1}{6}\sum_{i,j,m,l=1}^{7}\gamma^{\prime}_{ml}\tr(M_{\gamma_{1}}\mathbb{O}_{i}\mathbb{O}_{l})\tr(M_{\gamma_{4}}\mathbb{O}_{j}\mathbb{O}_{m})\tr(M_{\gamma_{2}}\mathbb{O}_{i})\tr(M_{\gamma_{3}}\mathbb{O}_{j})\nonumber\\
&+\frac{1}{2}\sum_{i,j,k=1}^{7}\alpha^{\prime\prime}_{kj}\tr(M_{\gamma_{2}\circ\gamma_{3}}\mathbb{O}_{i}\mathbb{O}_{k})\tr(M_{\gamma_{1}}\mathbb{O}_{i})\tr(M_{\gamma_{4}}\mathbb{O}_{j})\nonumber\\
&+\frac{1}{2}\sum_{i,j,k=1}^{7}\beta^{\prime\prime}_{kj}\tr(M_{\gamma_{2}\circ\gamma_{3}^{-1}}\mathbb{O}_{i}\mathbb{O}_{k})\tr(M_{\gamma_{1}}\mathbb{O}_{i})\tr(M_{\gamma_{4}}\mathbb{O}_{j})\nonumber\\
&+\frac{1}{6}\sum_{i,j,m,l=1}^{7}\gamma^{\prime\prime}_{ml}\tr(M_{\gamma_{2}}\mathbb{O}_{i}\mathbb{O}_{l})\tr(M_{\gamma_{3}}\mathbb{O}_{j}\mathbb{O}_{m})\tr(M_{\gamma_{1}}\mathbb{O}_{i})\tr(M_{\gamma_{4}}\mathbb{O}_{j})\nonumber\\
&+\frac{1}{2}\sum_{i,j,k=1}^{7}\alpha^{\prime\prime\prime}_{kj}\tr(M_{\gamma_{2}\circ\gamma_{4}}\mathbb{O}_{i}\mathbb{O}_{k})\tr(M_{\gamma_{1}}\mathbb{O}_{i})\tr(M_{\gamma_{3}}\mathbb{O}_{j})\nonumber\\
&+\frac{1}{2}\sum_{i,j,k=1}^{7}\beta^{\prime\prime\prime}_{kj}\tr(M_{\gamma_{2}\circ\gamma_{4}^{-1}}\mathbb{O}_{i}\mathbb{O}_{k})\tr(M_{\gamma_{1}}\mathbb{O}_{i})\tr(M_{\gamma_{3}}\mathbb{O}_{j})\nonumber\\
&+\frac{1}{6}\sum_{i,j,m,l=1}^{7}\gamma^{\prime\prime\prime}_{ml}\tr(M_{\gamma_{2}}\mathbb{O}_{i}\mathbb{O}_{l})\tr(M_{\gamma_{4}}\mathbb{O}_{j}\mathbb{O}_{m})\tr(M_{\gamma_{1}}\mathbb{O}_{i})\tr(M_{\gamma_{3}}\mathbb{O}_{j})\nonumber,
\end{align}
where each of $\alpha$, $\beta$, $\gamma$, $\alpha^{\prime}$, $\beta^{\prime}$, $\gamma^{\prime}$, $\alpha^{\prime\prime}$, $\beta^{\prime\prime}$, $\gamma^{\prime\prime}$, $\alpha^{\prime\prime\prime}$, $\beta^{\prime\prime\prime}$ and $\gamma^{\prime\prime\prime}$ is in the fundamental representation of $G_2$. The $G_2$-gauge invariant observables similar to summands of the types (\ref{exmpl-gauge-invrnt-obsrvbl-1}) and (\ref{exmpl-gauge-invrnt-obsrvbl-2}) correspond to $r=1$, $n_{1}=2$, $s=n_{2}=0$ and $t=1$ while the $(0,1)$-matrix $K$ reads as$\begin{bmatrix}1&1\\\end{bmatrix}$ in theorem \ref{G2-gauge-inv-obsrvbl}. Exotic $G_2$-gauge invariant observables similar to summands of the type (\ref{exmpl-gauge-invrnt-obsrvbl-3}), on the other hand, correspond to $r=n_{1}=2$, $s=0$, $n_{2}=1$, $t=2$ and the $(0,1)$-matrices $K$ and $Q$ are given by 
$$K=Q=\begin{bmatrix}1&0\\0&1\end{bmatrix}$$ in theorem \ref{G2-gauge-inv-obsrvbl}.

\section{Conclusion and outlook}
In this paper, we have generalized the Goldman bracket to the case of $G_{2}$ gauge group. The expression for the Poisson bracket between traces of $G_{2}$-valued monodromy matrices reveals the existence of a gauge invariant term of new type which were not present in the cases of classical gauge groups obtained by Goldman in \cite{goldmant}. An infinite set of such exotic $G_2$-gauge inavriant observables is obtained that is closed under Poisson bracket. As a by-product, we present an alternative derivation of the well-known Goldman's bracket for the following gauge groups: $GL(n,\mathbb{R})$, $U(n)$, $SL(n,\mathbb{R})$, $SU(n)$, $Sp(2n,\mathbb{R})$ and $SO(n)$. In future, we plan to extend our formalism to find Goldman-type brackets for the other exceptional gauge groups: $F_{4}$, $E_{6}$, $E_{7}$ and $E_{8}$. It would also be interesting to see how the quantum Goldman bracket turns out to be for the $G_{2}$ gauge group and for the other exceptional groups in comparison with the one computed in (p 428, \cite{quantized-gldmn-qg}) for $SL(2,\mathbb{R})$. We plan to investigate the underlying quantum geometry for the exceptional gauge groups by looking at the relevant quantum Goldman bracket in a future publication.

\section{Acknowledgements}
The author would like to thank the referee for many fruitful suggestions in order to improve the quality of the presentation of the manuscript. He would also like to thank Prof. Dmitry Korotkin for proposing the problem to him and gratefully acknowledges many fruitful discussions with him during the preparation of the manuscript. The author acknowledges a grant from National Natural Science Foundation of China (NSFC) under Grant No. 11550110186.

\section{Appendix}\label{sec-app}
This Appendix is devoted to rather lengthy and boring proofs of some of the results quoted in the paper.
\medskip

\prf {\bf of proposition \ref{prop-casimir-g2}}
\medskip

Let us, first, concentrate on one of the commuting pairs $\{C_{1},C_{9}\}$ of the $\mathfrak{g}_{2}$ basis elements (see (\ref{basis-elements-g2})). We have,
\begin{eqnarray}\label{first-pair-first}
-(C_{1}\otimes C_{1})&=&-\frac{1}{2}[(-e_{47}-e_{56}+e_{65}+e_{74})\otimes(-e_{47}-e_{56}+e_{65}+e_{74})]\nonumber\\
&=&-\frac{1}{2}(e_{47}\otimes e_{47}+e_{56}\otimes e_{56}+e_{65}\otimes e_{65}+e_{74}\otimes e_{74})\nonumber\\
&+&\frac{1}{2}(e_{47}\otimes e_{74}+e_{56}\otimes e_{65}+e_{65}\otimes e_{56}+e_{74}\otimes e_{47})\nonumber\\
&-&\frac{1}{2}(e_{47}\otimes e_{56}-e_{47}\otimes e_{65}+e_{56}\otimes e_{47}-e_{56}\otimes e_{74}\nonumber\\
&-&e_{65}\otimes e_{47}+e_{65}\otimes e_{74}-e_{74}\otimes e_{56}+e_{74}\otimes e_{65}).
\end{eqnarray}
We also have,
\begin{eqnarray}\label{first-pair-second}
-(C_{9}\otimes C_{9})&=&-\frac{1}{6}[(-2e_{12}+2e_{21}+e_{47}-e_{56}+e_{65}-e_{74})\nonumber\\
&&\;\;\otimes(-2e_{12}+2e_{21}+e_{47}-e_{56}+e_{65}-e_{74})]\nonumber\\
&=&-\frac{2}{3}(e_{12}\otimes e_{12}+e_{21}\otimes e_{21})+\frac{2}{3}(e_{12}\otimes e_{21}+e_{21}\otimes e_{12})\nonumber\\
&&+\frac{1}{6}(e_{47}\otimes e_{74}+e_{56}\otimes e_{65}+e_{65}\otimes e_{56}+e_{74}\otimes e_{47})\nonumber\\
&&-\frac{1}{6}(e_{47}\otimes e_{47}+e_{56}\otimes e_{56}+e_{65}\otimes e_{65}+e_{74}\otimes e_{74})\nonumber\\
&&+\frac{1}{3}(e_{12}\otimes e_{47}-e_{12}\otimes e_{56}+e_{12}\otimes e_{65}-e_{12}\otimes e_{74}\nonumber\\
&&-e_{21}\otimes e_{47}+e_{21}\otimes e_{56}-e_{21}\otimes e_{65}+e_{21}\otimes e_{74}\nonumber\\
&&+e_{47}\otimes e_{12}-e_{47}\otimes e_{21}+e_{56}\otimes e_{21}+e_{65}\otimes e_{12}\nonumber\\
&&- e_{65}\otimes e_{21}-e_{74}\otimes e_{12}+e_{74}\otimes e_{21}-e_{56}\otimes e_{12})\nonumber\\
&&+\frac{1}{6}(e_{47}\otimes e_{56}-e_{47}\otimes e_{65}+e_{56}\otimes e_{47}-e_{56}\otimes e_{74}\nonumber\\
&&-e_{65}\otimes e_{47}+e_{65}\otimes e_{74}-e_{74}\otimes e_{56}+e_{74}\otimes e_{65}).
\end{eqnarray}
Adding (\ref{first-pair-first}) to (\ref{first-pair-second}), one obtains the following
\begin{eqnarray}\label{first-pair}
\lefteqn{-(C_{1}\otimes C_{1})-(C_{9}\otimes C_{9})}\nonumber\\
&&=-\frac{2}{3}(e_{12}\otimes e_{12}+e_{21}\otimes e_{21}+e_{47}\otimes e_{47}+e_{56}\otimes e_{56}+e_{65}\otimes e_{65}+e_{74}\otimes e_{74})\nonumber\\
&&\;\;\;\;+\frac{2}{3}(e_{12}\otimes e_{21}+e_{21}\otimes e_{12}+e_{47}\otimes e_{74}+e_{56}\otimes e_{65}+e_{65}\otimes e_{56}+e_{74}\otimes e_{47})\nonumber\\
&&\;\;\;\;-\frac{1}{3}(e_{47}\otimes e_{56}-e_{47}\otimes e_{65}+e_{56}\otimes e_{47}-e_{56}\otimes e_{74}-e_{65}\otimes e_{47}+e_{65}\otimes e_{74}\nonumber\\
&&\;\;\;\;-e_{74}\otimes e_{56}+e_{74}\otimes e_{65}-e_{12}\otimes e_{47}+e_{12}\otimes e_{56}-e_{12}\otimes e_{65}+e_{12}\otimes e_{74}\nonumber\\
&&\;\;\;\;+e_{21}\otimes e_{47}-e_{21}\otimes e_{56}+e_{21}\otimes e_{65}-e_{21}\otimes e_{74}-e_{47}\otimes e_{12}+e_{47}\otimes e_{21}\nonumber\\
&&\;\;\;\;-e_{56}\otimes e_{21}+e_{56}\otimes e_{12}-e_{65}\otimes e_{12}+e_{65}\otimes e_{21}+e_{74}\otimes e_{12}-e_{74}\otimes e_{21})\nonumber\\
&&=-\frac{2}{3}(e_{12}\otimes e_{12}+e_{21}\otimes e_{21}+e_{47}\otimes e_{47}+e_{56}\otimes e_{56}+e_{65}\otimes e_{65}+e_{74}\otimes e_{74})\nonumber\\
&&\;\;\;\;+\frac{2}{3}(e_{12}\otimes e_{21}+e_{21}\otimes e_{12}+e_{47}\otimes e_{74}+e_{56}\otimes e_{65}+e_{65}\otimes e_{56}+e_{74}\otimes e_{47})\nonumber\\
&&\;\;\;\;-\frac{1}{3}[e_{47}\otimes(e_{56}-e_{65}+e_{21}-e_{12})+e_{74}\otimes(e_{65}-e_{56}+e_{12}-e_{21})\nonumber\\
&&\;\;\;\;\;\;\;\;\;+e_{56}\otimes(e_{47}-e_{74}+e_{12}-e_{21})+e_{65}\otimes(e_{74}-e_{47}+e_{21}-e_{12})\nonumber\\
&&\;\;\;\;\;\;\;\;\;+e_{12}\otimes(e_{56}-e_{47}-e_{65}+e_{74})+e_{21}\otimes(e_{47}-e_{56}+e_{65}-e_{74})]\nonumber\\
&&=-\frac{2}{3}(e_{12}\otimes e_{12}+e_{21}\otimes e_{21}+e_{47}\otimes e_{47}+e_{56}\otimes e_{56}+e_{65}\otimes e_{65}+e_{74}\otimes e_{74})\nonumber\\
&&\;\;\;\;+\frac{2}{3}(e_{12}\otimes e_{21}+e_{21}\otimes e_{12}+e_{47}\otimes e_{74}+e_{56}\otimes e_{65}+e_{65}\otimes e_{56}+e_{74}\otimes e_{47})\nonumber\\
&&\;\;\;\;+\frac{1}{3}[(e_{12}-e_{21})\otimes(e_{12}-e_{21}-e_{56}-e_{74}+e_{65}+e_{47})\nonumber\\
&&\;\;\;\;\;\;\;+(e_{47}-e_{74})\otimes(e_{47}-e_{74}-e_{56}-e_{21}+e_{65}+e_{12})\nonumber\\
&&\;\;\;\;\;\;\;+(-e_{56}+e_{65})\otimes(e_{65}-e_{56}+e_{47}+e_{12}-e_{74}-e_{21})]\nonumber\\
&&\;\;\;\;-\frac{1}{3}(e_{12}-e_{21})\otimes(e_{12}-e_{21})-\frac{1}{3}(e_{47}-e_{74})\otimes(e_{47}-e_{74})\nonumber\\
&&\;\;\;\;-\frac{1}{3}(e_{65}-e_{56})\otimes(e_{65}-e_{56})\nonumber\\
&&=-\frac{2}{3}(e_{12}\otimes e_{12}+e_{21}\otimes e_{21}+e_{47}\otimes e_{47}+e_{56}\otimes e_{56}+e_{65}\otimes e_{65}+e_{74}\otimes e_{74})\nonumber\\
&&\;\;\;\;+\frac{2}{3}(e_{12}\otimes e_{21}+e_{21}\otimes e_{12}+e_{47}\otimes e_{74}+e_{56}\otimes e_{65}+e_{65}\otimes e_{56}+e_{74}\otimes e_{47})\nonumber\\
&&\;\;\;\;+\frac{1}{3}(e_{12}-e_{21}+e_{47}-e_{74}+e_{65}-e_{56})\otimes(e_{12}-e_{21}+e_{47}-e_{74}+e_{65}-e_{56})\nonumber\\
&&\;\;\;\;-\frac{1}{3}(e_{12}\otimes e_{12}+e_{21}\otimes e_{21})+\frac{1}{3}(e_{12}\otimes e_{21}+e_{21}\otimes e_{12})\nonumber\\
&&\;\;\;\;-\frac{1}{3}(e_{47}\otimes e_{47}+e_{74}\otimes e_{74})+\frac{1}{3}(e_{47}\otimes e_{74}+e_{74}\otimes e_{47})\nonumber\\
&&\;\;\;\;-\frac{1}{3}(e_{65}\otimes e_{65}+e_{56}\otimes e_{56})+\frac{1}{3}(e_{65}\otimes e_{56}+e_{56}\otimes e_{65})\nonumber\\
&&=-(e_{12}\otimes e_{12}+e_{21}\otimes e_{21}+e_{47}\otimes e_{47}+e_{74}\otimes e_{74}+e_{56}\otimes e_{56}+e_{65}\otimes e_{65})\nonumber\\
&&\;\;\;\;+(e_{12}\otimes e_{21}+e_{21}\otimes e_{12}+e_{47}\otimes e_{74}+e_{74}\otimes e_{47}+e_{56}\otimes e_{65}+e_{65}\otimes e_{56})\nonumber\\
&&\;\;\;\;+\frac{1}{3}(e_{12}-e_{21}+e_{47}-e_{74}+e_{65}-e_{56})\otimes(e_{12}-e_{21}+e_{47}-e_{74}+e_{65}-e_{56}).\nonumber\\
\end{eqnarray}

Similarly, for the commuting pairs of the basis elements $\{C_{2},C_{10}\}$, $\{C_{3},C_{8}\}$, $\{C_{4},C_{11}\}$, $\{C_{5},C_{12}\}$, $\{C_{6},C_{13}\}$ and $\{C_{7},C_{14}\}$,  one obtains
\begin{eqnarray}\label{second-pair}
\lefteqn{-(C_{2}\otimes C_{2})-(C_{10}\otimes C_{10})}\nonumber\\
&&=-(e_{13}\otimes e_{13}+e_{31}\otimes e_{31}+e_{46}\otimes e_{46}+e_{64}\otimes e_{64}+e_{57}\otimes e_{57}+e_{75}\otimes e_{75})\nonumber\\
&&\;\;\;\;+(e_{13}\otimes e_{31}+e_{31}\otimes e_{13}+e_{46}\otimes e_{64}+e_{64}\otimes e_{46}+e_{57}\otimes e_{75}+e_{75}\otimes e_{57})\nonumber\\
&&\;\;\;\;+\frac{1}{3}(e_{31}-e_{13}+e_{46}-e_{64}+e_{57}-e_{75})\otimes(e_{31}-e_{13}+e_{46}-e_{64}+e_{57}-e_{75}).\nonumber\\
\end{eqnarray}
\begin{eqnarray}\label{third-pair}
\lefteqn{-(C_{3}\otimes C_{3})-(C_{8}\otimes C_{8})}\nonumber\\
&&=-(e_{23}\otimes e_{23}+e_{32}\otimes e_{32}+e_{45}\otimes e_{45}+e_{54}\otimes e_{54}+e_{67}\otimes e_{67}+e_{76}\otimes e_{76})\nonumber\\
&&\;\;\;\;+(e_{23}\otimes e_{32}+e_{32}\otimes e_{23}+e_{45}\otimes e_{54}+e_{54}\otimes e_{45}+e_{67}\otimes e_{76}+e_{76}\otimes e_{67})\nonumber\\
&&\;\;\;\;+\frac{1}{3}(-e_{32}+e_{23}-e_{54}+e_{45}-e_{67}+e_{76})\otimes(-e_{32}+e_{23}-e_{54}+e_{45}-e_{67}+e_{76}).\nonumber\\
\end{eqnarray}
\begin{eqnarray}\label{fourth-pair}
\lefteqn{-(C_{4}\otimes C_{4})-(C_{11}\otimes C_{11})}\nonumber\\
&&=-(e_{14}\otimes e_{14}+e_{41}\otimes e_{41}+e_{27}\otimes e_{27}+e_{72}\otimes e_{72}+e_{36}\otimes e_{36}+e_{63}\otimes e_{63})\nonumber\\
&&\;\;\;\;+(e_{14}\otimes e_{41}+e_{41}\otimes e_{14}+e_{27}\otimes e_{72}+e_{72}\otimes e_{27}+e_{36}\otimes e_{63}+e_{63}\otimes e_{36})\nonumber\\
&&\;\;\;\;+\frac{1}{3}(e_{14}-e_{41}+e_{36}-e_{63}+e_{72}-e_{27})\otimes(e_{14}-e_{41}+e_{36}-e_{63}+e_{72}-e_{27}).\nonumber\\
\end{eqnarray}
\begin{eqnarray}\label{fifth-pair}
\lefteqn{-(C_{5}\otimes C_{5})-(C_{12}\otimes C_{12})}\nonumber\\
&&=-(e_{15}\otimes e_{15}+e_{51}\otimes e_{51}+e_{26}\otimes e_{26}+e_{62}\otimes e_{62}+e_{37}\otimes e_{37}+e_{73}\otimes e_{73})\nonumber\\
&&\;\;\;\;+(e_{15}\otimes e_{51}+e_{51}\otimes e_{15}+e_{26}\otimes e_{62}+e_{62}\otimes e_{26}+e_{37}\otimes e_{73}+e_{73}\otimes e_{37})\nonumber\\
&&\;\;\;\;+\frac{1}{3}(e_{51}-e_{15}+e_{62}-e_{26}+e_{73}-e_{37})\otimes(e_{51}-e_{15}+e_{62}-e_{26}+e_{73}-e_{37}).\nonumber\\
\end{eqnarray}
\begin{eqnarray}\label{sixth-pair}
\lefteqn{-(C_{6}\otimes C_{6})-(C_{13}\otimes C_{13})}\nonumber\\
&&=-(e_{16}\otimes e_{16}+e_{61}\otimes e_{61}+e_{25}\otimes e_{25}+e_{52}\otimes e_{52}+e_{34}\otimes e_{34}+e_{43}\otimes e_{43})\nonumber\\
&&\;\;\;\;+(e_{16}\otimes e_{61}+e_{61}\otimes e_{16}+e_{25}\otimes e_{52}+e_{52}\otimes e_{25}+e_{34}\otimes e_{43}+e_{43}\otimes e_{34})\nonumber\\
&&\;\;\;\;+\frac{1}{3}(e_{61}-e_{16}+e_{25}-e_{52}+e_{34}-e_{43})\otimes(e_{61}-e_{16}+e_{25}-e_{52}+e_{34}-e_{43}).\nonumber\\
\end{eqnarray}
\begin{eqnarray}\label{seventh-pair}
\lefteqn{-(C_{7}\otimes C_{7})-(C_{14}\otimes C_{14})}\nonumber\\
&&=-(e_{17}\otimes e_{17}+e_{71}\otimes e_{71}+e_{24}\otimes e_{24}+e_{42}\otimes e_{42}+e_{35}\otimes e_{35}+e_{53}\otimes e_{53})\nonumber\\
&&\;\;\;\;+(e_{17}\otimes e_{71}+e_{71}\otimes e_{17}+e_{24}\otimes e_{42}+e_{42}\otimes e_{24}+e_{35}\otimes e_{53}+e_{53}\otimes e_{35})\nonumber\\
&&\;\;\;\;+\frac{1}{3}(e_{17}-e_{71}+e_{24}-e_{42}+e_{53}-e_{35})\otimes(e_{17}-e_{71}+e_{24}-e_{42}+e_{53}-e_{35}).\nonumber\\
\end{eqnarray}

Adding (\ref{first-pair}), (\ref{second-pair}), (\ref{third-pair}), (\ref{fourth-pair}), (\ref{fifth-pair}) and (\ref{sixth-pair}) to (\ref{seventh-pair}), one obtains
\begin{eqnarray}\label{casimir-g2-final-exprssn}
\Gamma&=&\sum_{i=1}^{7}-(C_{i}\otimes C_{i})\nonumber\\
&=&\sum_{i\neq j} e_{ij}\otimes e_{ji}-\sum_{i\neq j} e_{ij}\otimes e_{ij}+\frac{1}{3}\sum_{i=1}^{7}\mathbb{O}_{i}\otimes \mathbb{O}_{i}\nonumber\\
&=&\left(\sum_{i\neq j} e_{ij}\otimes e_{ji}+\sum_{i=1}^{7} e_{ii}\otimes e_{ii}\right)-\left(\sum_{i\neq j} e_{ij}\otimes e_{ij}+\sum_{i=1}^{7}e_{ii}\otimes e_{ii}\right)+\frac{1}{3}\sum_{i=1}^{7}\mathbb{O}_{i}\otimes \mathbb{O}_{i}\nonumber\\
&=&\sum_{i,j=1}^{7}e_{ij}\otimes e_{ji}-\sum_{i,j
=1}^{7}e_{ij}\otimes e_{ij}+\frac{1}{3}\sum_{i=1}^{7}\mathbb{O}_{i}\otimes \mathbb{O}_{i}.
\end{eqnarray}
\qed

\prf {\bf of theorem \ref{G2-gauge-inv-obsrvbl}}
\medskip

Without loss of generality, we assume $t=2$, i.e., we will choose the $(0,1)$-matrices $K$ and $Q$ to be of order $2\times n_{1}$ and $2\times(2n_{2}-s)$, respectively, with the entries in the second row of the matrix $K$ and in that of the $t\times(2n_{2}-2s)$-block of $Q$ to be all zeros. One then obtains for any $g\in G_2$ in its 7-dimensional fundamental representation
\begin{align}
&\sum\limits_{l_{j}\in\{1,2,...,7\}}\tr (gM_{\gamma_1}g^{-1}\mathbb{O}_{l_1})...\tr(gM_{\gamma_{r}}g^{-1}\mathbb{O}_{l_r})\tr(gM_{\gamma_{r+1}}g^{-1}\mathbb{O}_{l_{r+1}})...\tr(gM_{\gamma_{n_{1}}}g^{-1}\mathbb{O}_{l_{n_{1}}})\nonumber\\
&\times\tr\left(gM_{\gamma_{n_{1}+1}}g^{-1}\mathbb{O}_{l_{1}}...\mathbb{O}_{l_{r}}\mathbb{O}_{l_{n_{1}+1}}...\mathbb{O}_{l_{2n_{1}-r}}\mathbb{O}_{l_{2n_{1}-r+1}}...\mathbb{O}_{l_{2n_{1}-r+s}}\mathbb{O}_{l_{2n_{1}-r+s+1}}\right.\nonumber\\
&
\left.\hspace{2.5in}\mathbb{O}_{l_{2n_{1}-r+n_{2}+1}}...\mathbb{O}_{l_{2n_{1}-r+n_{2}}}\mathbb{O}_{l_{2n_{1}+2n_{2}-r-s}}\right)\nonumber\\
&\times\tr(gM_{\gamma_{n_{1}+2}}g^{-1}\mathbb{O}_{l_{2n_{1}-r+1}}...\mathbb{O}_{l_{2n_{1}-r+s}})\nonumber\\
&\times (g\alpha^{1}g^{-1})_{l_{r+1},l_{n_{1}+1}}...(g\alpha^{n_{1}-r}g^{-1})_{l_{n_{1}},l_{2n_{1}-r}}\nonumber\\
&\times (g\beta^{1}g^{-1})_{l_{2n_{1}-r+s+1},l_{2n_{1}-r+n_{2}+1}}...(g\beta^{n_{2}-s}g^{-1})_{l_{2n_{1}-r+n_{2}},l_{2n_{1}+2n_{2}-r-s}}\nonumber\\
&=\sum\limits_{l_{j},m_{j},k_{j},p_{j}\in\{1,2,...,7\}}\tr (M_{\gamma_1}\mathbb{O}_{k_1})...\tr(M_{\gamma_r}\mathbb{O}_{k_r})\tr(M_{\gamma_{r+1}}\mathbb{O}_{k_{r+1}})...\tr(M_{\gamma_{n_{1}}}\mathbb{O}_{k_{n_{1}}})\nonumber\\
&\times\tr\left(M_{\gamma_{n_{1}+1}}\mathbb{O}_{m_{1}}\mathbb{O}_{m_{2}}...\mathbb{O}_{m_{r}}\mathbb{O}_{k_{n_{1}+1}}...\mathbb{O}_{k_{2n_{1}-r}}\mathbb{O}_{k_{2n_{1}-r+1}}...\mathbb{O}_{k_{2n_{1}-r+s}}\mathbb{O}_{k_{2n_{1}-r+s+1}}\right.\nonumber\\
&\left.\hspace{2.5in}\mathbb{O}_{k_{2n_{1}-r+n_{2}+1}}...\mathbb{O}_{k_{2n_{1}-r+n_{2}}}\mathbb{O}_{k_{2n_{1}+2n_{2}-r-s}}\right)\nonumber\\
&\times\tr\left(M_{\gamma_{n_{1}+2}}\mathbb{O}_{p_{2n_{1}-r+1}}...\mathbb{O}_{p_{2n_{1}-r+s}}\right)\nonumber\\
&\times g^{-1}_{k_{1},l_{1}}...g^{-1}_{k_{r},l_{r}}g^{-1}_{k_{r+1},l_{r+1}}...g^{-1}_{k_{n_{1}},l_{n_{1}}}g^{-1}_{m_{1},l_{1}}...g^{-1}_{m_{r},l_{r}}g^{-1}_{k_{n_{1}+1},l_{n_{1}+1}}...g^{-1}_{k_{2n_{1}-r},l_{2n_{1}-r}}\nonumber\\
&\times g^{-1}_{k_{2n_{1}-r+1},l_{2n_{1}-r+1}}...g^{-1}_{k_{2n_{1}-r+s},l_{2n_{1}-r+s}}g^{-1}_{k_{2n_{1}-r+s+1},l_{2n_{1}-r+s+1}}g^{-1}_{k_{2n_{1}-r+n_{2}+1},l_{2n_{1}-r+n_{2}+1}}\nonumber\\
&\times... g^{-1}_{k_{2n_{1}-r+n_{2}},l_{2n_{1}-r+n_{2}}}g^{-1}_{k_{2n_{1}+2n_{2}-r-s},l_{2n_{1}+2n_{2}-r-s}}g^{-1}_{p_{2n_{1}-r+1},l_{2n_{1}-r+1}}...g^{-1}_{p_{2n_{1}-r+s},l_{2n_{1}-r+s}}\nonumber\\
&\times (g\alpha^{1}g^{-1})_{l_{r+1},l_{n_{1}+1}}...(g\alpha^{n_{1}-r}g^{-1})_{l_{n_{1}},l_{2n_{1}-r}}\nonumber\\
&\times (g\beta^{1}g^{-1})_{l_{2n_{1}-r+s+1},l_{2n_{1}-r+n_{2}+1}}...(g\beta^{n_{2}-s}g^{-1})_{l_{2n_{1}-r+n_{2}},l_{2n_{1}+2n_{2}-r-s}}\nonumber\\
&=\sum\limits_{l_{j},m_{j},k_{j},p_{j}\in\{1,2,...,7\}}\tr (M_{\gamma_1}\mathbb{O}_{k_1})...\tr(M_{\gamma_r}\mathbb{O}_{k_r})\tr(M_{\gamma_{r+1}}\mathbb{O}_{k_{r+1}})...\tr(M_{\gamma_{n_{1}}}\mathbb{O}_{k_{n_{1}}})\nonumber\\
&\times\tr\left(M_{\gamma_{n_{1}+1}}\mathbb{O}_{m_{1}}\mathbb{O}_{m_{2}}...\mathbb{O}_{m_{r}}\mathbb{O}_{k_{n_{1}+1}}...\mathbb{O}_{k_{2n_{1}-r}}\mathbb{O}_{k_{2n_{1}-r+1}}...\mathbb{O}_{k_{2n_{1}-r+s}}\mathbb{O}_{k_{2n_{1}-r+s+1}}\right.\nonumber\\
&\left.\hspace{2.5in}\mathbb{O}_{k_{2n_{1}-r+n_{2}+1}}...\mathbb{O}_{k_{2n_{1}-r+n_{2}}}\mathbb{O}_{k_{2n_{1}+2n_{2}-r-s}}\right)\nonumber\\
&\times\tr\left(M_{\gamma_{n_{1}+2}}\mathbb{O}_{p_{2n_{1}-r+1}}...\mathbb{O}_{p_{2n_{1}-r+s}}\right)\nonumber\\
&\times\delta_{m_{1},k_{1}}...\delta_{m_{r},k_{r}}\delta_{k_{2n_{1}-r+1},p_{2n_{1}-r+1}}...\delta_{k_{2n_{1}-r+s},p_{2n_{1}-r+s}}(g^{-1}_{k_{r+1},l_{r+1}})(g\alpha^{1}g^{-1})_{l_{r+1},l_{n_{1}+1}}\nonumber\\
&\times(g_{l_{n_{1}+1},k_{n_{1}+1}})...(g^{-1}_{k_{n_{1}},l_{n_{1}}})(g\alpha^{n_{1}-r}g^{-1})_{l_{n_{1}},l_{2n_{1}-r}}(g_{l_{2n_{1}-r},k_{2n_{1}-r}})\nonumber\\
&\times(g^{-1}_{k_{2n_{1}-r+s+1},l_{2n_{1}-r+s+1}})(g\beta^{1}g^{-1})_{l_{2n_{1}-r+s+1},l_{2n_{1}-r+n_{2}+1}}(g_{l_{2n_{1}-r+n_{2}+1},k_{2n_{1}-r+n_{2}+1}}).....\nonumber\\
&\times(g^{-1}_{k_{2n_{1}-r+n_{2}},l_{2n_{1}-r+n_{2}}})(g\beta^{n_{2}-s}g^{-1})_{l_{2n_{1}-r+n_{2}},l_{2n_{1}+2n_{2}-r-s}}(g_{l_{2n_{1}+2n_{2}-r-s},k_{2n_{1}+2n_{2}-r-s}})\nonumber\\
&=\sum\limits_{k_{j}\in\{1,2,...,7\}}\tr(M_{\gamma_1}\mathbb{O}_{k_1})...\tr(M_{\gamma_r}\mathbb{O}_{k_r})\tr(M_{\gamma_{r+1}}\mathbb{O}_{k_{r+1}})...\tr(M_{\gamma_{n_{1}}}\mathbb{O}_{k_{n_{1}}})\nonumber\\
&\times\tr\left(M_{\gamma_{n_{1}+1}}\mathbb{O}_{k_{1}}\mathbb{O}_{k_{2}}...\mathbb{O}_{k_{r}}\mathbb{O}_{k_{n_{1}+1}}...\mathbb{O}_{k_{2n_{1}-r}}\mathbb{O}_{k_{2n_{1}-r+1}}...\mathbb{O}_{k_{2n_{1}-r+s}}\mathbb{O}_{k_{2n_{1}-r+s+1}}\right.\nonumber\\
&\left.\hspace{2.5in}\mathbb{O}_{k_{2n_{1}-r+n_{2}+1}}...\mathbb{O}_{k_{2n_{1}-r+n_{2}}}\mathbb{O}_{k_{2n_{1}+2n_{2}-r-s}}\right)\nonumber\\
&\times\tr(M_{\gamma_{n_{1}+2}}\mathbb{O}_{k_{2n_{1}-r+1}}...\mathbb{O}_{k_{2n_{1}-r+s}})\nonumber\\
&\times (\alpha^{1})_{k_{r+1},k_{n_{1}+1}}...(\alpha^{n_{1}-r})_{k_{n_{1}},k_{2n_{1}-r}}\nonumber\\
&\times (\beta^{1})_{k_{2n_{1}-r+s+1},k_{2n_{1}-r+n_{2}+1}}...(\beta^{n_{2}-s})_{k_{2n_{1}-r+n_{2}},k_{2n_{1}+2n_{2}-r-s}},
\end{align}
where we have repeatedly applied lemma \ref{lemma-transfrng-left-to-right-act} in order to go to the second line from the first one. We have also made successive applications of the fact that $g_{ij}$ is orthogonal during the course of the proof.
\qed.


\begin{thebibliography}{99}


\bibitem{M-theory-g2-holonomy-physics}
 B. S. Acharya,
\newblock {\em M theory, $G_{2}$-manifolds and four-dimensional physics,}
\newblock Class. Quant. Grav., {\bf 19} (2002), 5619-5653.





\bibitem{CSref}
Alexandrov, S., Geiller, M. and Noui, K.,
\newblock {\em Spin Foams and Canonical Quantization,}
\newblock SIGMA, {\bf 8} 055 (2012).




\bibitem{M-theory-g2-holonomy}
 M. Atiyah and E. Witten,
\newblock {\em M-theory dynamics on a manifold of $G(2)$
holonomy,}
\newblock Adv. Theor. Math. Phys., {\bf 6} (2003), 1-106.




\bibitem{exceptional-holonomy}
 Robert L. Bryant,
\newblock {\em Metrics with exceptional holonomy,}
\newblock Annals of Mathematics, {\bf 126} (1987), 525-576.



\bibitem{eulerangle}
 S. L. Cacciatori, B. L. Cerchiai, A. Della Vedova, G.
Ortenzi, and A. Scotti,
\newblock {\em Euler angles for $G_{2}$ ,}
\newblock J. Math. Phys., {\bf 46}, 083512 (2005).



\bibitem{cattaneo}
A. Cattaneo, J. Fr\"ohlich, B. Pedrini,
\newblock {\em Topological field theory interpretation
of string topology,}
\newblock Commun. Math. Phys., {\bf 240}, 397--421 (2003).





\bibitem{chas-sull}
M. Chas and D. Sullivan,
\newblock {\em String toplogy,}
\newblock preprint (1999), arXiv: math.GT/9911159.



\bibitem{faddeevetalt}
L. D. Faddeev and L. A. Takhtajan,
\newblock {\em Hamiltonian Methods in the Theory of Solitons,}
\newblock Springer, Berlin, 1987.



\bibitem{goldmant}
Goldman, W.,
\newblock {\em Invariant functions on Lie groups and Hamiltonian flows
of surface group representations,}
\newblock Invent. math., {\bf 85}, 263--302 (1986).




\bibitem{color-screening}
J. Greensite, K. Langfeld, \v{S}. Olejnik, H. Reinhardt, and T. Tok,
\newblock {\em Color Screening, Casimir Scaling, and Domain Structure in $G(2)$ and $SU(N)$ Gauge Theories,}
\newblock Phys.Rev. D, {\bf 75} (2007), 034501.




\bibitem{excptnl-confinement}
K. Holland, P. Minkowski, M. Pepe and U.-J. Wiese,
\newblock {\em Exceptional confinement in $G(2)$ gauge theory,}
\newblock Nucl.Phys. B, {\bf 668} (2003), 207--236.


\bibitem{mathoverflow}
Jos\'{e} Figueroa-O'Farrill,
\newblock{{\em Answer to Mathoverflow question}, \url{http://mathoverflow.net/a/214533/78032}.}




\bibitem{octonionic-rep}
S. De Leo and K. Abdel-Khalek,
\newblock {\em Octonionic representations of $GL(8,\mathbb{R})$ and $GL(4,\mathbb{C})$,}
\newblock J. Math. Phys., {\bf 38}, 582--598 (1997).





\bibitem{quantized-gldmn}
J.E. Nelson and R.F. Picken,
\newblock {\em Constant connections, quantum
holonomies and the Goldman
bracket,}
\newblock Adv. Theor. Math. Phys., {\bf 9}, 407--433 (2005).




\bibitem{quantized-gldmn-qg}
J.E. Nelson and R.F. Picken,
\newblock {\em A quantum Goldman bracket in (2 + 1) quantum
gravity,}
\newblock J. Phys. A: Math. Theor, {\bf 41} (2008), 304011 (9pp).




\bibitem{excptnl-deconfinement}
M. Pepe and U.-J. Wiese,
\newblock {\em Exceptional Deconfinement in $G(2)$ Gauge Theory,}
\newblock Nucl.Phys. B, {\bf 768} (2007), 21-37 .





\bibitem{sen-zwiebach}
A. Sen and B. Zwiebach,
\newblock {\em Background independent algebraic structures in closed string field theory,}
\newblock Commun. Math. Phys., {\bf 177}, 305--326 (1996).





\bibitem{turaev}
Turaev, V. G.,
\newblock {\em Skein quantization of Poisson algebras of loops on surfaces,}
\newblock Ann. Sci. Ecole Norm. Sup. (4) , {\bf 24,} No. 6, (1991), 635--704 .




\bibitem{g2-gluodynamics}
Bj\"{o}rn H. Wellegehausen, Andreas Wipf, and Christian Wozar,
\newblock {\em Casimir Scaling and String Breaking in $G_{2}$ Gluodynamics,}
\newblock Phys.Rev. D, {\bf 83} (2011), 016001.




\bibitem{wittencmpt}
Witten, E.,
\newblock {\em Quantum field theory and the Jones polynomial,}
\newblock Commun. Math. Phys., {\bf 121}, 351--399 (1989).




\bibitem{wittenbrst}
Witten, E.,
\newblock {\em Topological sigma models,}
\newblock Commun. Math. Phys., {\bf 118}, 411--449 (1988).



\bibitem{wolpert}
Wolpert, S.,
\newblock {\em On the Symplectic Geometry of Deformations of Hyperbolic Surfaces,}
\newblock Ann. Math., {\bf 117}(1983), 207--234.



\end{thebibliography}
\end{document}